\documentclass[vecphys]{svmult}

\usepackage{makeidx}         % allows index generation
\usepackage{graphicx}        % standard LaTeX graphics tool
\usepackage{multicol}        % used for the two-column index
\usepackage[bottom]{footmisc}% places footnotes at page bottom
\makeindex             % used for the subject index
\begin{document}

\title*{The SARG Planet Search}
\author{S. Desidera \inst{1}\and
        R. Gratton  \inst{1}\and
        M. Endl     \inst{2}\and
        A.F. Martinez Fiorenzano  \inst{3}\and
        M. Barbieri \inst{4} \and
        R. Claudi   \inst{1}\and
        R. Cosentino \inst{3,5}\and
        S. Scuderi \inst{5} \and
        M. Bonavita \inst{1,6}}

\authorrunning{S. Desidera et al.}
\institute{INAF - Osservatorio Astronomico di Padova, Italy
\texttt{silvano.desidera@oapd.inaf.it}
\and McDonald Observatory, The University of Texas at Austin, Austin, USA
\and INAF - Fundacion Galileo Galilei, Santa Cruz de La Palma, Spain
\and LAM - Observatoire de Marseille, France
\and INAF - Osservatorio Astrofisico di Catania, Italy
\and Dip. di Astronomia, Universit\'a di Padova, Italy}

\maketitle

\section{Introduction}
\label{s:intro}

The search for planets in multiple systems allows to improve our knowledge
on planet formation and evolution. On one hand, the frequency of planets
in binary systems has a strong effect on the global frequency of planets,
as more than half of solar type stars are in binary or multiple systems 
(\cite{duq91}).
On the other hand, the properties of planets in binaries, and any
difference with those of the planets orbiting single stars would
shed light on the effects caused by the presence of the companions.

The search for planets in binaries can follow two 
complementary approaches.
The first one is to perform dedicated surveys looking for planets in 
binary systems.
Several programs  currently in progress focusing on different types of
binaries are described in this book.
In this chapter, we describe the first planet search entirely 
dedicated to binary systems, the  survey on-going at TNG using
the high resolution spectrograph SARG.

The second approach is to study the binarity of the hosts of planets discovered
in general surveys, which include many binary stars in their lists in spite
of some selection biases against them. 
Indeed, the first analysis on the properties of planets in binaries
showed the occurrence of some differences with respect to those
orbiting single stars (\cite{zucker02}, \cite{egg04}).

In Sect.~\ref{s:bin} we summarize our recent work on the statistical properties
of planets in binaries.
In Sect.~\ref{s:abu} we present the second major science goal of the
SARG survey, the search for abundance anomalies caused by the
ingestion of planetary material by the central star. 
In Sections \ref{s:sample} to \ref{s:limits} we present the sample, 
the observing and analysis
priocedures, and the preliminary results of the SARG planet search.
Finally, in Sect.~\ref{s:freq} we present some preliminary conclusions
on the frequency of planets in binary systems.

\section{Properties of planets in binary systems}
\label{s:bin}

More than 40 planets have been found in binary or multiple systems
An updated compilation was recently assembled by some of us (\cite{db06}).
We performed a  statistical analysis of the properties of planets in binaries
and the comparison with respect to those orbiting single stars,
based on the planet and stellar parameters listed in the
Catalog of Nearby Exoplanets by \cite{butler06}.

\begin{figure}
\centering
\includegraphics[height=6.0cm]{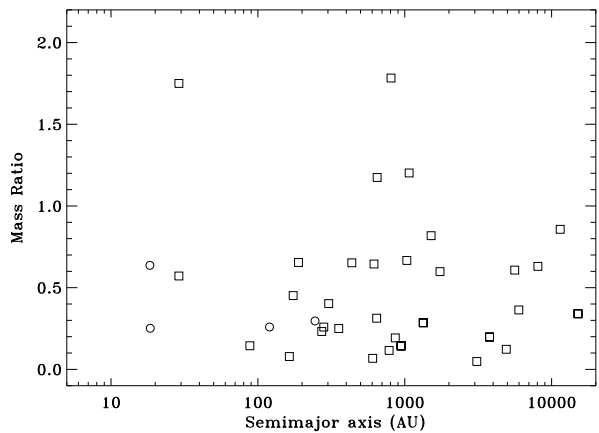}
\caption{Mass ratio vs semimajor axis of the binary orbit  for 
         stars with planets in binary systems. 
         Open circles represent
         the pairs for which binary orbit is available, open squares
         the pairs for which only the binary separation is available.
         From \cite{db06}.}
\label{f:planets_bin}     
\end{figure}

Fig.~\ref{f:planets_bin} shows the mass ratio vs semimajor axis  for 
stars with planets in multiple systems. For hierarchical triple
systems in which the planet orbits the isolated companion, the masses 
of the binary companions to the planet host 
are summed.
It results that planets might exist in binaries with very different
properties. In some cases (e.g. very low mass companions at
a projected separation larger than 1000 AU) the dynamical effects
of the companion on the formation and evolution of the planetary system
might be very limited, while in the cases of very tight binaries
the presence  of the planet represents a challenge for the
current models of planet formation (\cite{hatzes05}).

\begin{figure}
\centering
\includegraphics[height=6.0cm]{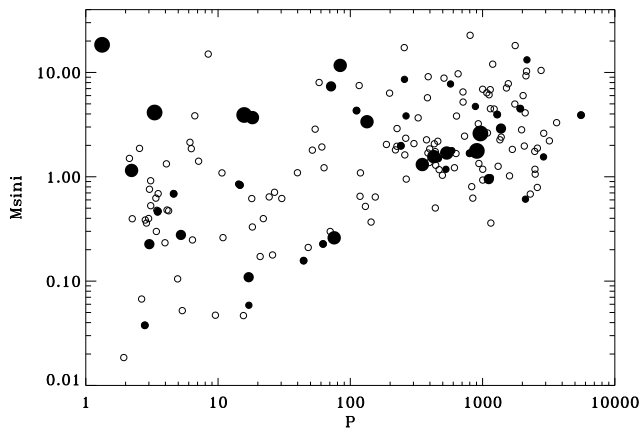}
\caption{Projected mass vs orbital period of extrasolar planets.
               Open circles: single stars; filled circles: binary stars.
               The size of the symbol is proportional to the critical 
               semimajor axis
               for dynamical stability (larger symbols refer to the 
               tighter binaries). From \cite{db06}.}
\label{f:msini_p}     
\end{figure}

To consider the effects of dynamical perturbation by the
stellar companion(s) we used the critical semiaxis for dynamical
stability of the planet $a_{crit}$  
\cite{holman}.
We choose $a_{crit}$ as a reference value because it is a physical 
quantity that  represents the 
dynamical effects due to a companion on planet formation and stability, 
including both the 
orbital parameters and mass ratio.

The critical semimajor axis $a_{crit}$ was used to divide the sample according
to the relevance of the dynamical effects. 
We define as 'tight' binaries those with $a_{crit}<75$~AU
and 'wide' binaries those with $a_{crit}>75$~AU. The limit corresponds to
a projected separation of about 200-300 AU depending on the mass ratio.

The statistical comparison to test the hypothesis that the
parameters of planets (mass, period, eccentricity) in tight and wide binaries 
and in single stars
can be drawn from the same parent distribution was performed 
using the Kolmogorov-Smirnov  test and the Mann-Whitney U 
test.

\begin{figure}
\centering
\includegraphics[height=4.1cm]{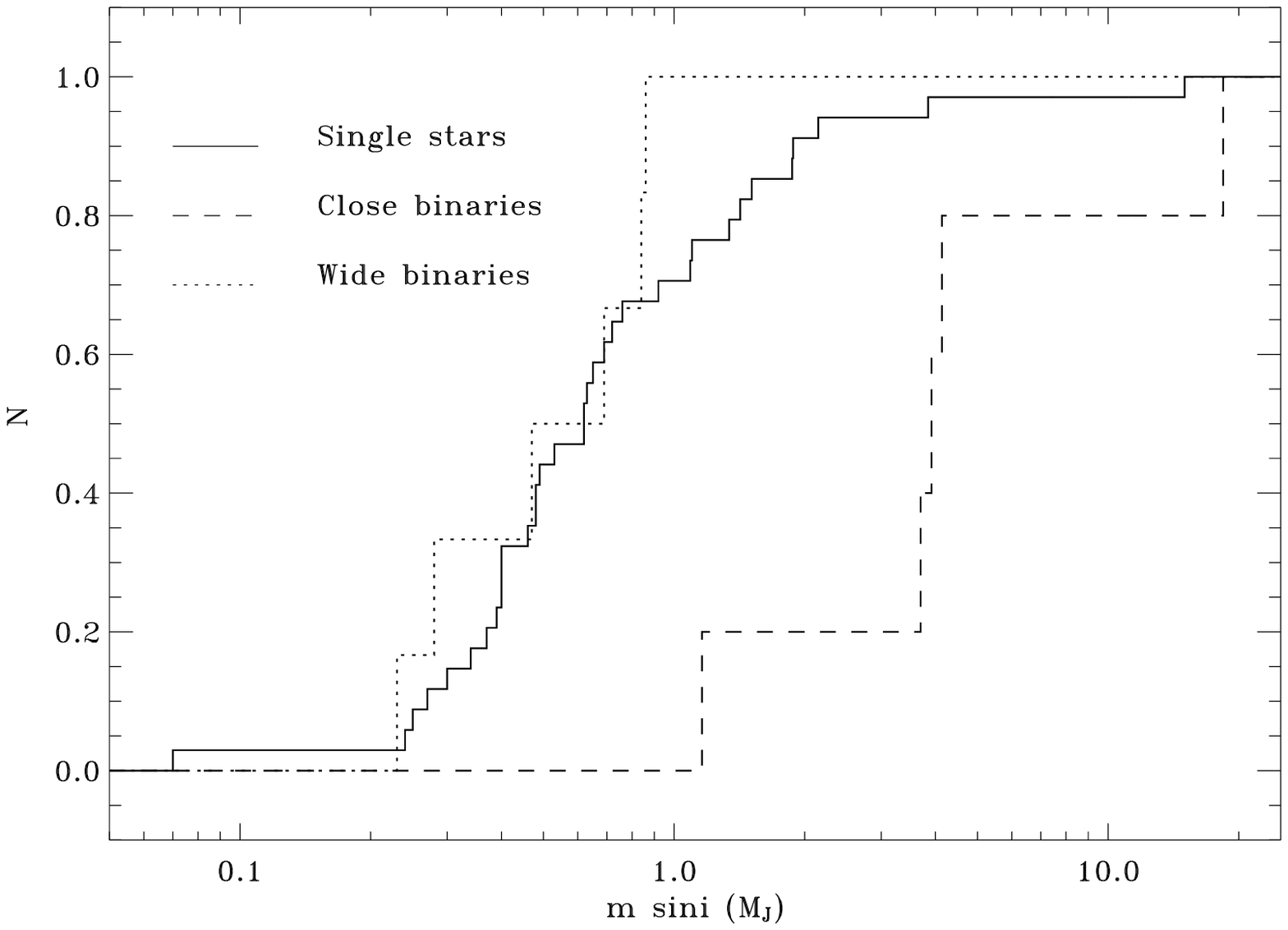}
\includegraphics[height=4.1cm]{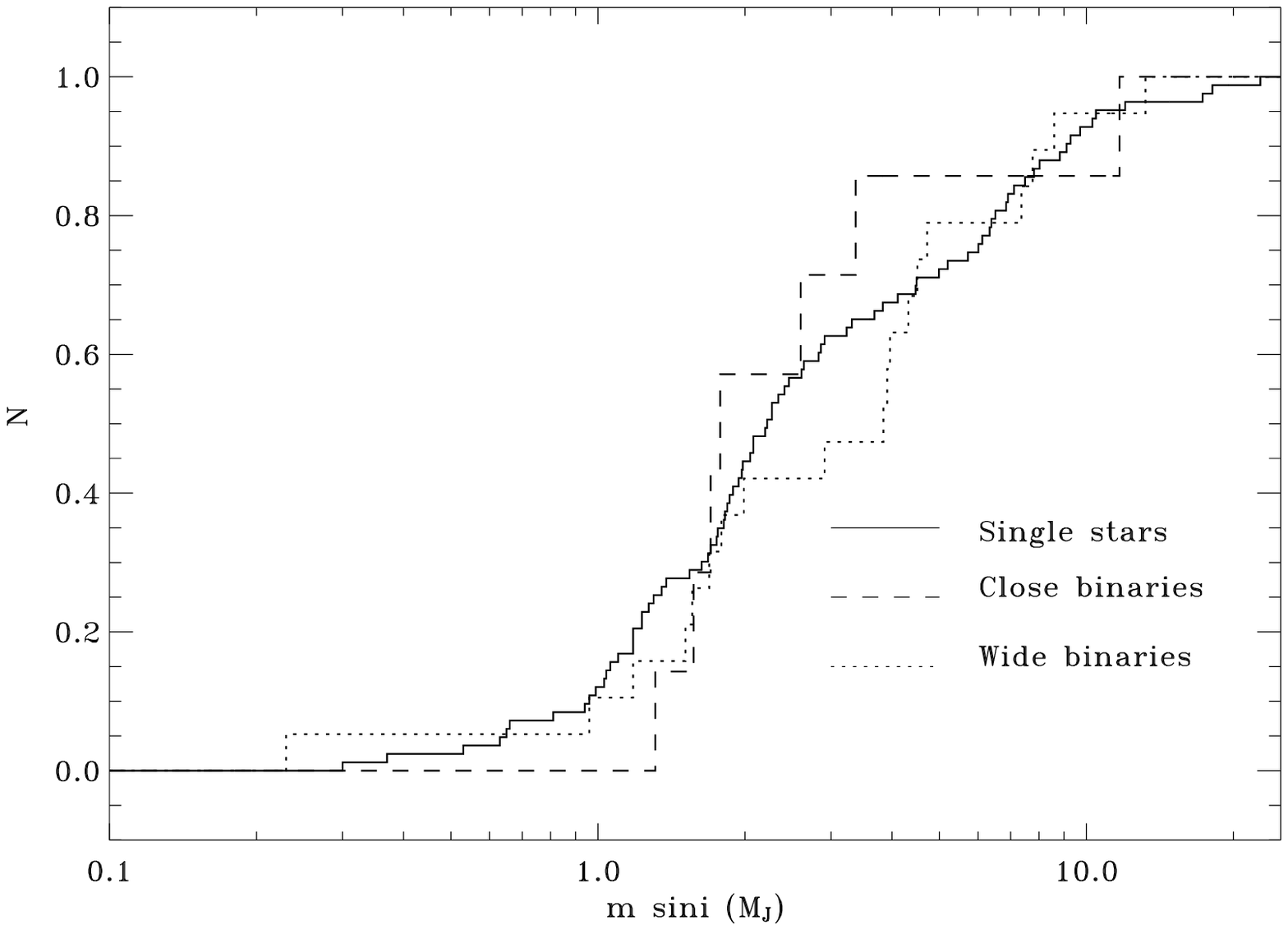}
\includegraphics[height=4.1cm]{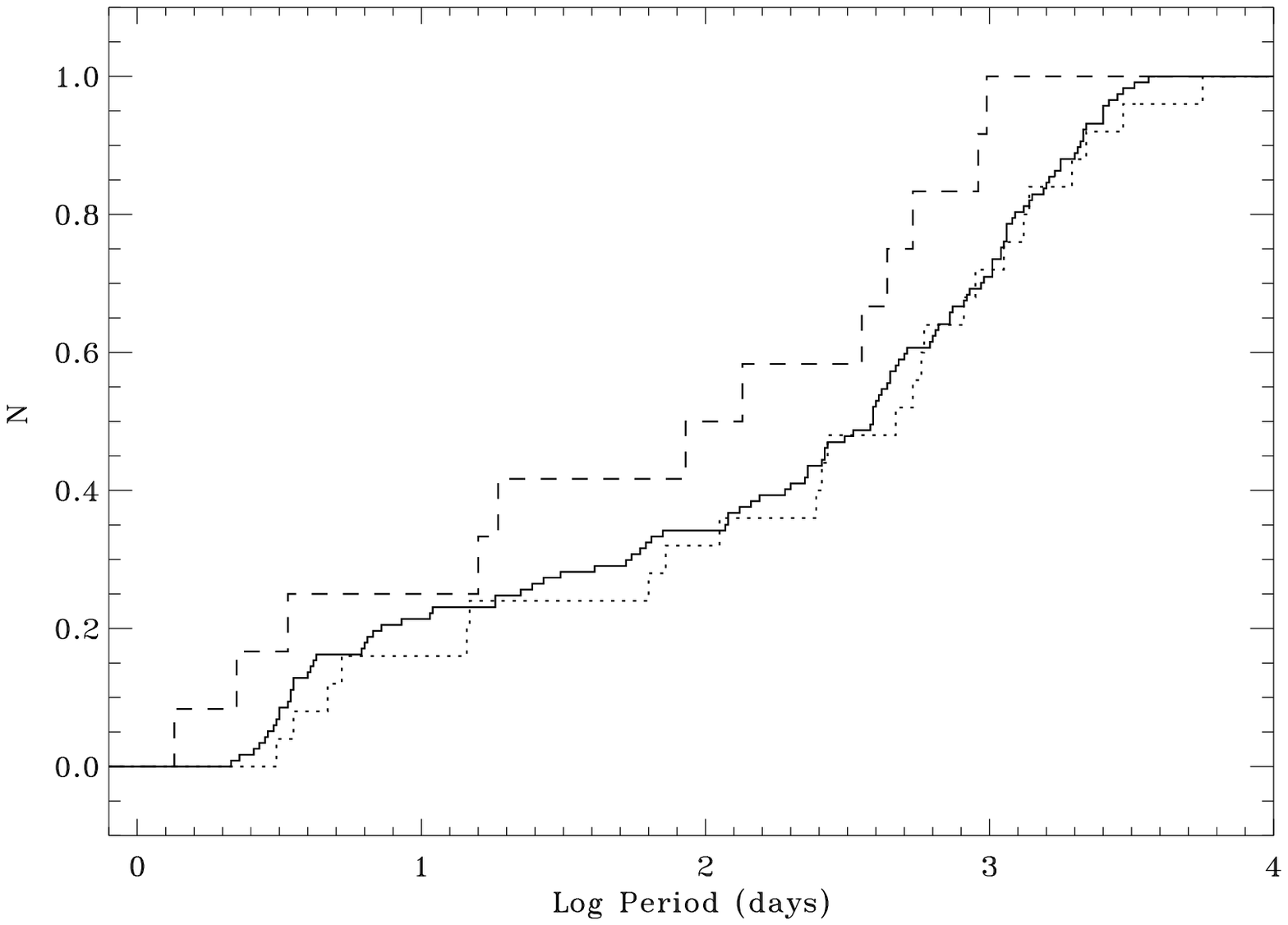}
\includegraphics[height=4.1cm]{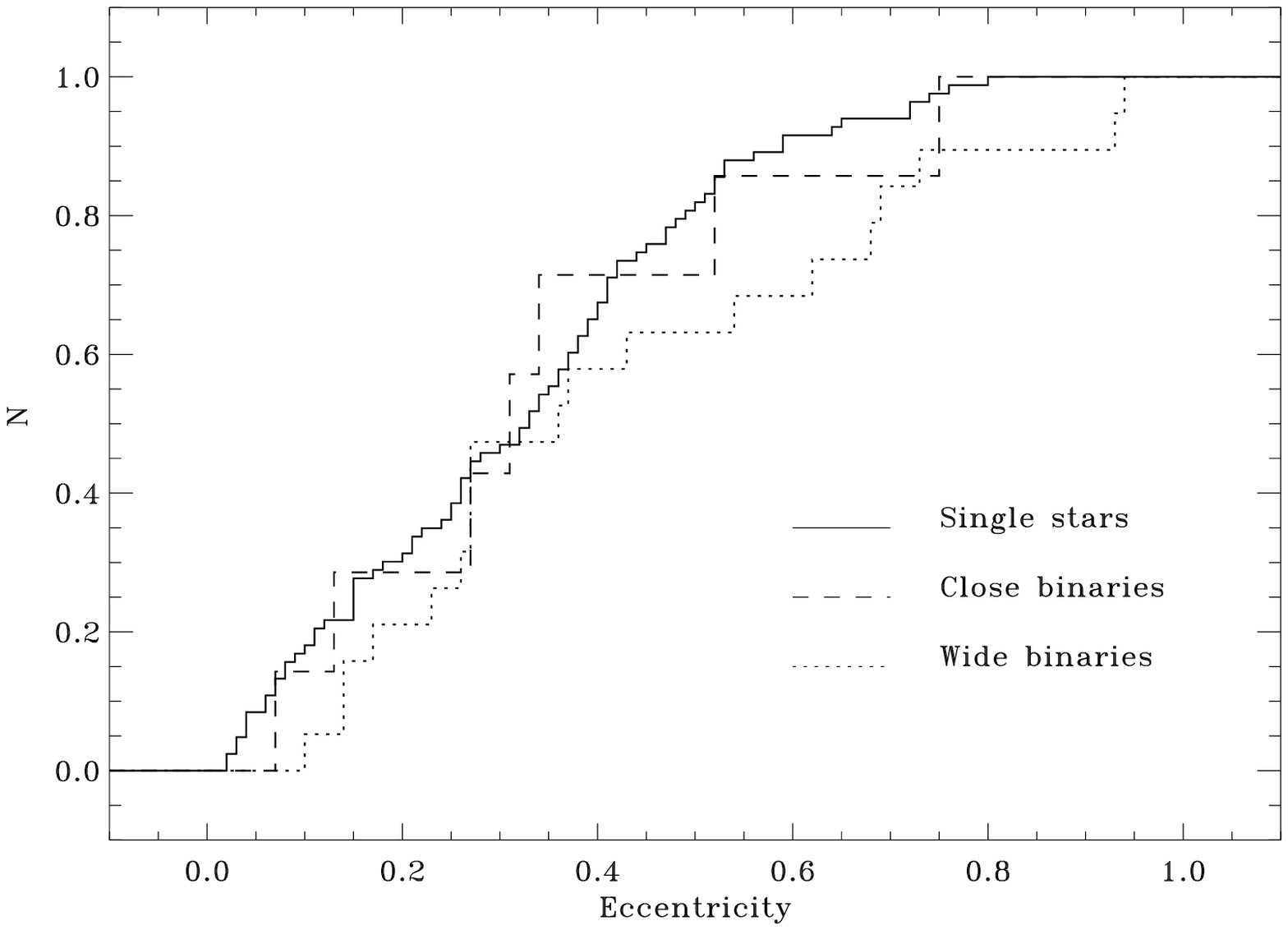}
\caption{Cumulative distributions of planets parameters for
planets orbiting single stars (continuous lines); components of wide 
binaries (dotted lines); components of tight binaries (dashed lines).
Upper left panel: mass distribution of planets with period shorter
than 40 days. Upper right panel: mass distribution of planets with 
period longer than 40 days. Lower left panel: period distribution. 
Lower right panel: eccentricity distribution of planets with 
period longer than 40 days.
Adapted from \cite{db06}.}
\label{f:cumul}     
\end{figure}

\begin{figure}
\centering
\includegraphics[height=6.0cm]{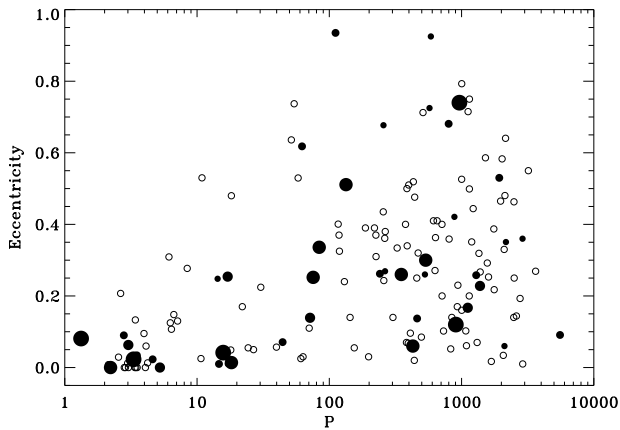}
\caption{Eccentricity vs orbital period for planets in 
               binaries (filled circles) and
               orbiting single stars (empty circles). 
               Different sizes of filled circles refer 
               to different periastron of the binary orbit 
               (larger sizes: closer orbits). From \cite{db06}.}
\label{f:p_e}     
\end{figure}

The following results were found (see Fig~\ref{f:cumul}):

\begin{itemize}
\item
The mass distribution of short period ($P<40$ days) planets 
in tight binaries is significantly ($>99$\%) different with respect
to that of planets orbiting single stars and components of wide
binaries. Massive, short period planets are mostly found 
in tight binaries (Fig.~\ref{f:msini_p}-\ref{f:cumul}).
This somewhat resembles
the evidence that short-period
spectroscopic binaries have in most cases a further companion (\cite{tok06}).
\item
The mass distributions of planets in wide orbits in tight and wide
binaries and in single stars are not significantly different.
\item
The differences in period distributions are  also not highly
significant.  
However, there is a marginal indication for a lack of long period
planets in tight binaries.
\item 
The eccentricity distribution of planets in tight binaries with 
periods longer than 40 days is not significantly 
different  to those
orbiting single stars.
On the other hand, there is a marginal indication for a larger eccentricity
of planets in wide binaries (Fig.~\ref{f:cumul}-\ref{f:p_e}).
\item
The occurrence of systems with more than one  planet 
around the components of wide binaries 
is similar with respect to that of planets 
orbiting single stars.
No multiple planets have been yet discovered instead around the 
components of tight binaries, but their small number 
makes the lack of multi-planet systems
not highly significant (probability of 15\% of occurring by chance).
\end{itemize}

We then conclude that planets in close binaries have different 
characteristics with respect
to those orbiting single stars and components of wide binaries.
The mass and period distributions of planets in wide binaries instead
are not statistically significant different 
to those of planets orbiting single stars.
The only marginally significant difference between planets orbiting single
stars and components of wide binaries concerns the planet eccentricity.
In any case, high planet eccentricities are not 
confined to planets in binaries,
and  the possible differences in eccentricity appears to be limited
to the range $e \ge 0.5-0.6$.
This indicates that there are mechanism(s) generating planet 
eccentricity up to 0.4-0.5 that are independent of the binarity
of the planet host, and are characteristic of formation and
evolution of a planetary system
(e.g. disk-planet interactions, planet-planet scattering).
These probably act
during or shortly after planet formation.
Further eccentricity enhancements, possibly linked to the presence
of a companion, might take place
at later epochs. 
In fact, \cite{takeda06} noted that most  
high-eccentricity planets orbit old stars (ages $>$5 Gyr).  
Mechanisms that require long time scales to modify 
planetary orbits, such as Kozai oscillations and  chaotic evolution
of planetary orbits induced by dynamical perturbations then seem favored.

These results indicate that a companion at large separation 
($\ge 500$~AU) probably does not  affect
too much the planet formation process around one of the components, while
the effects of the companions are much more relevant at small
separation, causing differences in the physical properties of the
planets.

The understanding of the formation mechanism of the planets in close
binaries is a key
problem. One possibility is that these planets formed before the binary
configuration was modifed by stellar encounters in the native star cluster
(\cite{pfhal06}). The alternative is that planets do form 
in close binaries in spite of the seemingly unfavourable conditions.
The exploration of the frequency and properties of planets at intermediate
binary separations (100-300 AU), the range of a large fraction
of the binaries of the SARG planet search, is  important to establish 
the separation required to show the peculiar features of planet properties.

\section{Binary systems as a tool to evidence the ingestion of planetary 
material by the central star}

\label{s:abu}

The evidence for a high metal content in stars harbouring 
planets is becoming stronger as planet discoveries cumulate
and suitable control samples are studied using strictly the same procedures
(\cite{santos04}, \cite{fv05}).
Two alternative hypotheses have been proposed to explain these observations:
either the high metallicity is responsible for the presence of planets,
making their formation easier; or the planets are the cause of
the high metallicity, because of pollution of metal-rich planetary material
onto the (outer region of the) central star (\cite{gonzalez}).

Infall of planetesimals on the star during the early phases of planet
formation is generally expected on the basis of current models
of planet formation.
The orbital migration proposed to explain the
occurrence of the close-in giant planets found by
radial velocity surveys also points to the infall on the star
of portions of the proto-planetary disk.

Most of the accretion is expected to take place
during the early phases of the evolution of the planetary system.
However, when a star is still in the phase of
gravitational contraction, its convective zone is much thicker than for 
main sequence stars (see e.g. \cite{murray}).
In this case, the metal-rich material should be uniformly distributed
by convective mixing over a large portion of the star, resulting in a
negligible photospheric chemical alteration even for rather large amounts
of accreted material.
Late accretion, when the star is approaching or has already reached the 
main sequence, is likely required to produce observable differences.
The ingestion of planets scattered toward the star by dynamical
interactions (\cite{marzari02}) might also
produce metallicity enhancements at late phases.

Murray and co-workers (\cite{murray}) found that the Sun
should have ingested some $2~M_{\oplus}$ of meteoritic material
(about $0.4~M_{\oplus}$ of iron) during its main-sequence lifetime, 
considering the drop of iron density 
in the asteroid region and the time distribution
of the impact craters. This corresponds to a metallicity 
enhancement of 0.017 dex.
Such a small abundance difference is not detectable when considering
a field star, for which no proper reference for the original unpolluted 
abundance is available.
In binary systems and star clusters instead such a reference is
provided by the other companion/members of the system.
Therefore, the  comparison of the chemical
composition of wide binaries is a very powerful approach to study
the occurrence of planetary pollution, provided that differential
abundance analysis with a precision of about 0.02 dex can be obtained.

If the high metallicity is the result of planets or planetesimal ingestion
(\cite{gonzalez}), some systematic difference is expected between members 
of a binary system  with and without planetary companions. 
On the other hand common metallicity between components should indicate 
a robust link between metallicity and formation process of planetary systems.

\section{The SARG sample}
\label{s:sample}

With the two science goals identified in Sections \ref{s:intro}-\ref{s:abu}, 
we started a few years ago
a radial velocity (RV) survey of the components of wide binaries.
We are using SARG, the high resolution spectrograph of the TNG
(\cite{sarg}), equipped with an iodine cell to derive high precision
RVs.

The sample was selected from the Hipparcos Multiple Star Catalog,
considering binaries in the magnitude range $7.0<V<10.0$, 
with magnitude difference between the components of $\Delta V < 1.0$,  
projected separation larger than 2 arcsec (to avoid contamination
of the spectra), parallax larger than 10 mas and error smaller
than 5 mas, with $B-V>0.45$ and spectral type later than F7.
About 50 pairs (100 stars) were selected.

\begin{figure}
\centering
\includegraphics[height=6.0cm]{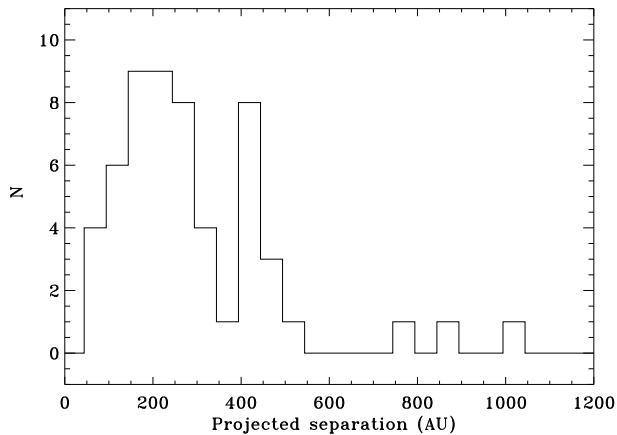}
\caption{Distribution of the projected separation in AU of the binaries
in the sample of the SARG survey.}
\label{f:rhoau}     
\end{figure}

\begin{figure}
\centering
\includegraphics[height=6.0cm]{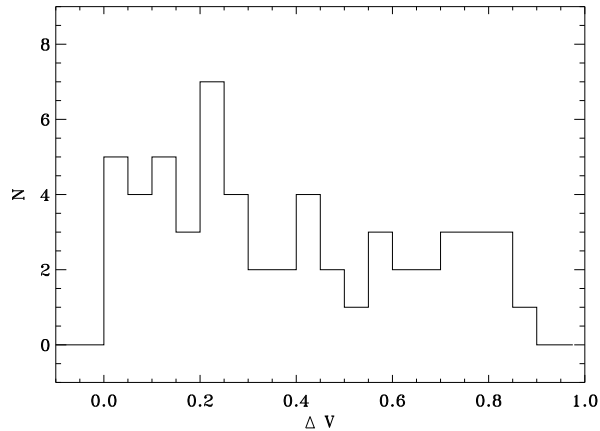}
\caption{Distribution of the visual magnitude difference in AU of the binaries
in the sample of the SARG survey.}
\label{f:deltav}     
\end{figure}

The sample is then formed by wide binaries with mass ratios close
to 1. Considering systems with similar components is crucial for the 
accuracy of the differential chemical
abundance analysis. 
Fig.~\ref{f:rhoau} shows the distribution of the projected separation
in AU. For most of the pairs, it results between 50 and 600 AU.
Fig.~\ref{f:deltav} shows the distribution of the V band magnitude difference
between the components.

\section{Observations}
\label{s:obs}

The observations used for the radial velocity determinations were
acquired with the SARG spectrograph \cite{sarg} 
using the Yellow Grism, that covers
the spectral range 4600-7900~\AA~without gaps, and the 0.25 arcsec slit.
The resulting resolution is  R=150000 (2 pixels sampling).
The iodine cell was inserted into the optical path, superimposing a dense 
forest of absorption lines used as reference spectrum for 
the radial velocity determination.
Exposure times were fixed in most cases at 15 minutes, 
to reduce the  errors in barycentric 
correction caused by the lack of knowledge of the exact flux mid time of the
exposure. 
A high signal to noise spectrum without the iodine cell
was also acquired for all the program stars, to be used for the
abundance analysis (see Sect.~\ref{s:chem}) and as template for
the radial velocity determination (see Sect.~\ref{s:rv}).

During the observations, the slit was usually oriented perpendicularly to
the separation of the components
to minimize the contamination of the spectra by the companion.
The closest pairs (separation 2-3 arcsec) were observed only in good seeing
conditions. In spite of these efforts, some residual contamination
of the spectra is present in a few cases.
This issue is discussed in Sect.~\ref{s:bis}.
The survey is in progress, up to now we have acquired on average
about 15 spectra per star.

\section{Abundance analysis}
\label{s:chem}

The abundance analysis of about half of the pairs of the SARG survey
was published in \cite{chem2} while in 
\cite{chem3} we studied 33 pairs of Southern declination
observed with the FEROS spectrograph at ESO-La Silla, selected
with similar criteria. Taking into account
the small overlap between the two samples, we have in hand 
the results for 50 pairs.

\begin{figure}
\centering
\includegraphics[height=6.0cm]{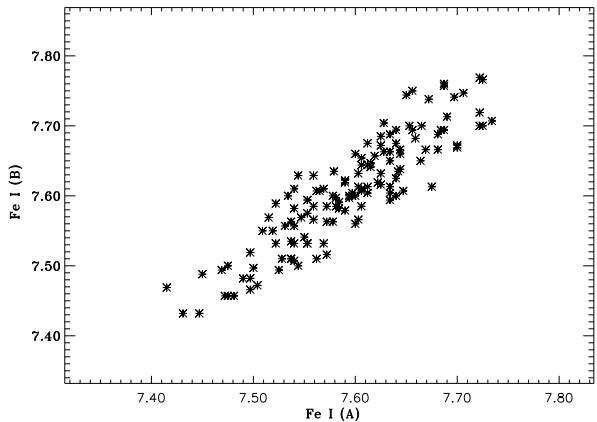}
\caption{Iron abundance derived for each line of the components of HIP 114914 A and B. 
A clear correlation is present, indicating that the use of a 
line-by-line differential analysis significantly reduces
the errors on abundance difference between the components.
From \cite{chem3}.}
\label{f:fe}      
\end{figure}

Performing a line-by-line differential analysis (Fig.~\ref{f:fe}) 
and  exploiting the physical link between the components (same distance
from the Sun), we found that errors in estimating the difference of iron
content between the two components of about 0.02 dex 
can be achieved for 
pairs with temperature differences
smaller than 300-400~K and slow-rotating components with effective 
temperatures in the range 5500-6300~K.
This is adequate for detailed  study
of chemical alterations in the external convective layer.

\begin{figure}
\centering
\includegraphics[height=6.0cm]{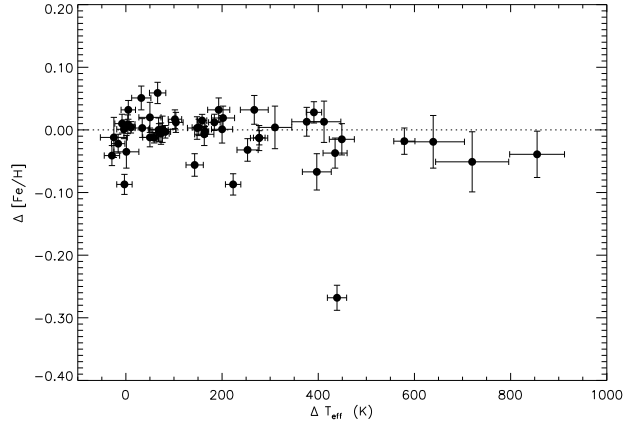}
\caption{Iron abundance difference between the components of pairs as a function of temperature difference for the pairs studied in \cite{chem2} and 
\cite{chem3}.}
\label{f:deltafe}      
\end{figure}

Most of the pairs have abundance difference smaller than 0.03 dex 
(Fig~\ref{f:deltafe}).
We found one case (HIP~64030=HD~113984) with a large (0.25 dex) abundance
difference. The primary of this binary appears to be a 
blue straggler, and  the abundance difference
might be due to the peculiar evolution  of the star (see Sect.~\ref{s:bs}).
A few other pairs show small abundance differences 
($\leq 0.09$ dex). In a few cases these differences suggest the
ingestion of a small amount of metal rich material, but
in others  they are likely spurious, because
of the large temperature difference between the components, the high level of
magnetic activity, that might cause alterations in the
stellar atmosphere or additional errors in our analysis because
of intrinsic variability, or possible contamination of the spectra by
an additional star in close orbit around one of the components.
Some cases of abundance differences involving 
pairs with warm ($T_{\rm eff} \geq 6000$ K) primaries might be due to 
the diffusion of heavy elements.

\begin{figure}
\centering
\includegraphics[height=6.0cm]{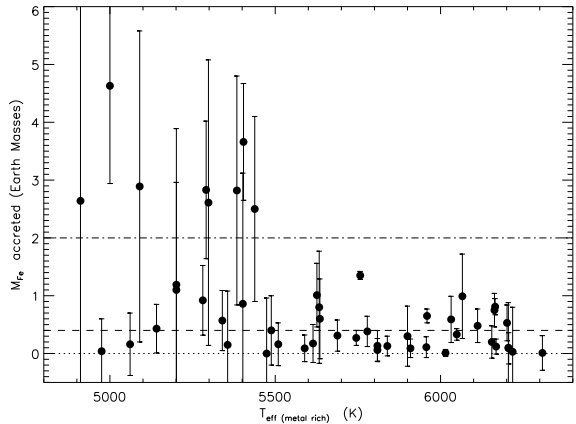}
\caption{Estimate of iron accreted by the metal-rich component of each pair 
as a function of its effective temperature, taking into account the mass of 
the mixing zone as in \cite{murray}.   %%% Murray et al.~(2001). 
The less severe limits at lower effective
temperatures are mostly due to the more massive convective zone of cool stars.
The horizontal lines show the amount of iron expected
to have been accreted by the Sun during the main sequence
lifetime ($0.4~M_{\oplus}$:  \cite{murray}),  
and the amount of iron corresponding to the upper limit
on abundance difference between the inner and outer
regions of the Sun according to helioseismology 
($2~M_{\oplus}$: \cite{winnick02}).
The mass of meteoritic material
is assumed to be about 5.5 times the mass of iron.
From \cite{chem3}.}

\label{f:accretion}       
\end{figure}

Fig.~\ref{f:accretion} shows the amount of iron accreted by the
nominally metal richer component to explain
the observed abundance difference.
For most of the slow-rotating stars warmer than 5500 K, characterized by a 
thinner convective envelope and for which our analysis appears to be of higher 
accuracy, this is similar to the estimates of rocky material accreted by the 
Sun during its main sequence lifetime (about 0.4 Earth masses of iron, 
\cite{murray}).
We then conclude that the occurrence of large alterations in stellar 
abundances 
caused by the ingestion  of metal rich, rocky material is not a common event.
For at least 65\% of the pairs with components warmer than 5500 K,
the limits on the amount of rocky material accreted
by the program stars are
comparable to the estimates of rocky material
accreted by the Sun during its main--sequence lifetime.

\subsection{The special case of the blue straggler HD 113984}
\label{s:bs}

The wide binary HIP64030=HD 113984 is the only pair in our sample
that shows a large (about $0.25$ dex) iron content difference.  
The positions of the components on the color magnitude diagram suggest
that the primary is a blue straggler.
Therefore, the abundance difference may be somewhat linked to the peculiar
evolutionary history of the system.

The analysis of additional elements beside iron (\cite{bs}) showed that the
abundance difference for the elements studied increases with 
increasing condensation temperature, suggesting that accretion of chemically 
fractionated material might have occurred in the system.
Alteration of C and N likely due to CNO processing
is also observed, as expected for the  mass transfer process occurring
during the formation of the blue straggler.
We also showed that the blue straggler component
is a spectroscopic binary with a period of 445 days
and moderate eccentricity, as typical for field blue stragglers 
(\cite{preston}).

Two scenarios were explored to explain the observed abundance pattern. 
In the first, all  abundance
anomalies arise on the blue straggler. 
If this is the case, the dust-gas separation may have been
occurred in a circumbinary disk around the blue straggler 
and its expected white dwarf companion, as observed in several 
RV Tauri and post AGB binaries \cite{vanwinckel}.
In the second scenario, accretion of dust-rich material
occurred on the secondary. This would also explain the anomalous 
carbon isotopic ratio of the secondary. 
Such a scenario requires that a substantial amount of mass
lost by the central binary has been accreted by the wide
component.

\begin{figure}
\centering
\includegraphics[height=6.0cm]{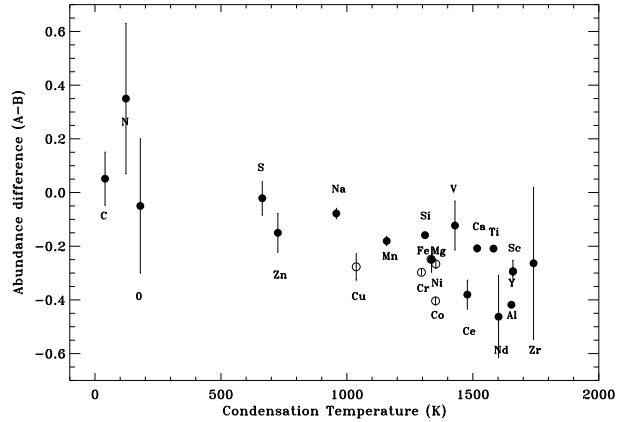}
\caption{Abundance difference for the components of HD 113984 as  a function
of the condensation temperature. From \cite{bs}}
\label{f:bs}       
\end{figure}

\subsection{Abundance difference between components for binary systems
with planetary companions}
\label{s:abuplanets}

The analysis of 50 pairs shown in Sect.~\ref{s:chem} suggests that 
the frequency of pairs with large alterations in chemical composition
is rather small. Therefore,
it seems unlikely that the ingestion of planetary material  can account for
the strong correlation between the frequency of planets and metallicity.

However, none of the  stars studied by \cite{chem2},\cite{chem3} are known
to host planets (most of the pairs of the FEROS sample are probably not
being searched for planets).
Therefore, it is interesting to consider the abundance difference
between the components of binary systems with/without planets.
We limit our analysis to pairs with similar components, as
errors in differential chemical abundances  
becomes larger for large temperature
difference (see discussion in \cite{chem3}).
 
Among the binary systems with planets, there are five pairs with mass
ratio between 0.8 and 1.2.
Only for 16 Cyg high-precision differential abundance analysis
between the components has been carried out.
Laws \& Gonzalez \cite{laws} found a small abundance difference
of 0.025 dex, with the planet-host (the secondary) being more metal-rich,
while \cite{takeda} did not confirm the reality of the small
abundance difference.

For the pairs HD 80606/7, HD 99491/2 and ADS 16402 
the standard abundance analysis
 does not reveal significant abundance difference
(see Table 1).
For HD 20781, the companion of the planet host HD 20782,
there are no high-resolution abundance analysis and the
abundance difference derived from Str\"omgren photometry
is not significant (errors about 0.1 dex).

\begin{table}
\begin{center}
\begin{tabular}{|c|c|c|l|}
\hline
 System  & Planet host & $\Delta$ [Fe/H]  & Ref. \\
\hline
16 Cyg     &   B  & $-0.025\pm0.009$ & \cite{laws} \\
16 Cyg     &   B  & $ 0.00\pm0.01$   & \cite{takeda} \\
HD 80606/7 &   A  & $-0.01\pm0.11$   & \cite{heiter} \\
HD 80606/7 &   A  & $+0.002\pm0.081$ & \cite{taylor05} \\
HD 99491/2 &   B  & $-0.02\pm0.03$   & \cite{vf05} \\
HD 99491/2 &   B  & $+0.04\pm0.13$   & \cite{heiter} \\
HD 99491/2 &   B  & $+0.076\pm0.059$ & \cite{taylor05}\\
HD 20781/2 &   A  & $+0.12\pm0.10$   & \cite{nordstrom} \\
ADS 16402  &   B  & $-0.01\pm0.05$   & \cite{bakos} \\

\hline
\end{tabular}
\caption{Abundance difference between the components of binary
planet hosts with similar components.}
\end{center}
\label{t:metal}
\end{table}

Summarizing, there are currently no evidence for large ($\ge 0.1$ dex) 
alterations of chemical abundances in the components of binary systems 
with/without planets. This supports the conclusion of our dedicated study
on the abundance difference between the components of binaries that
large alteration of chemical abundance caused by the ingestion of
planetary material are rare, if any.

\section{Radial velocities}
\label{s:rv}

\begin{figure}
\centering
\includegraphics[height=6.0cm]{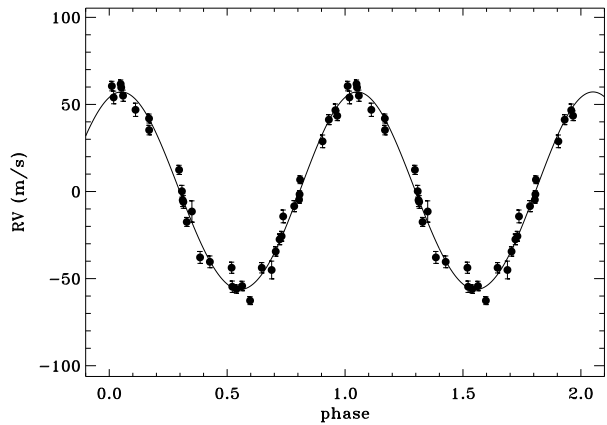}
\caption{Radial velocities of 51 Peg obtained with SARG phased to
the known orbital period.}
\label{f:51peg}  
\end{figure}

High precision 
RVs for the stars in the SARG sample 
were determined using the AUSTRAL code (\cite{austral})
as described in Desidera et al.~(\cite{hd219542}). 
On average we acquired up to now about 15 spectra per star.
Typical errors are 2-3 m/s for bright
stars observed as standards to monitor instrument performances 
(Fig.~\ref{f:51peg}) and 3-10 m/s for the $V\sim 7-9$ program stars.

\section{Planet candidates and low amplitude variables}

The RV time series are being searched for periodic variations 
as data cumulate.
No clear planet detection emerged up to now.
A couple of candidates have false alarm probabilities of about 1\%, but
they are of fairly low amplitude and further data are required
for confirmation.

Some further stars show RV variability above internal errors. 
In most cases this can be explained by stellar activity jitter and
residual contamination of the spectra from the companion (see
Sect.~\ref{s:bis}).

\begin{figure}
\centering
\includegraphics[height=6.0cm]{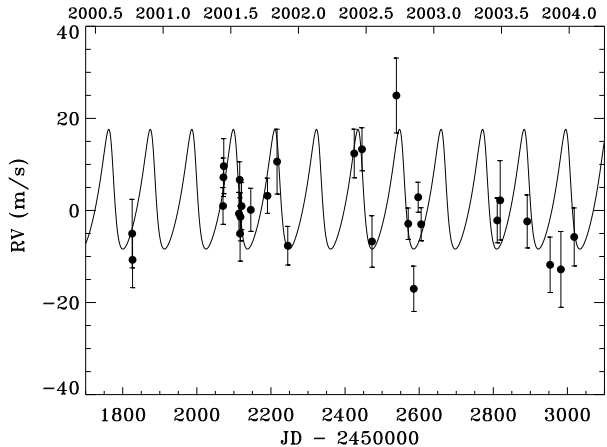}
\caption{Radial velocity curve for HD~219542~B. The data taken in
        the 2003 season do not follow the tentative orbital solution
        previously derived in \cite{hd219542} (overplotted as a solid line).
        From \cite{letter}. }
\label{f:hd219542b}  
\end{figure}

One case we investigated in detail is that of HD~219542B.
The 2000-2002 data indicated a possible periodicity of 111 days with a 
significance of about 97\% (\cite{hd219542}). However, the continuation of
the observations revealed that the RV variations are likely due to
stellar activity (see Fig.~\ref{f:hd219542b}; \cite{letter}).
In particular, the chromospheric emission measurements 
indicate that HD~219542~B underwent a phase of enhanced
stellar activity in 2002 while the activity level has been
lower in both 2001 and 2003.

\section{New triple systems and stars with long term trends}

\begin{figure}
\centering
\includegraphics[height=6.0cm]{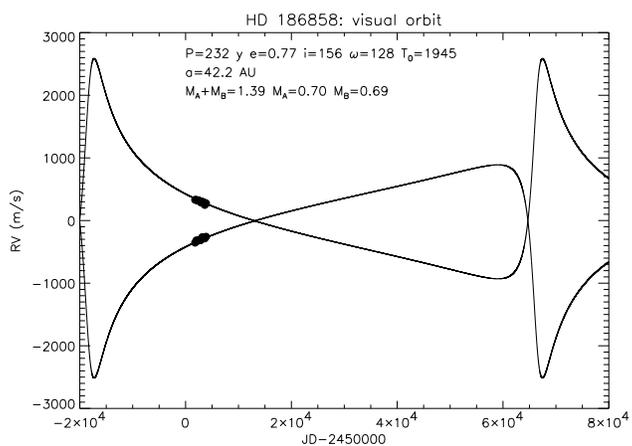}
\caption{Continuous lines: predicted RV curve for the components of the
binary system HD 186858 according to the visual+astrometric solution
derived by \cite{soder}. Filled circles: high-precision RV obtained with SARG
over 6 years. The RV slopes of each components and their absolute RV
difference follow very well the orbital solution. }
\label{f:hd186858}  
\end{figure}

More than 10\% of the stars in the sample show long term linear or nearly
linear trends. In a few cases the trends are due to the known companion,
as trends with opposite sign and nearly the same magnitude are observed
for the two components. Fig.~\ref{f:hd186858} shows the case of HD 186858,
for which a reliable visual+astrometric solution was presented by
\cite{soder}. The RV slopes of each components and their absolute RV
difference follow very well the orbital solution.
The full characterization of the binary orbit and individual masses
of the systems we are surveying is useful for the
study of the frequency binary systems with/without planets,
as described in Sect.~\ref{s:freq}.

\begin{figure}
\centering
\includegraphics[height=6.0cm]{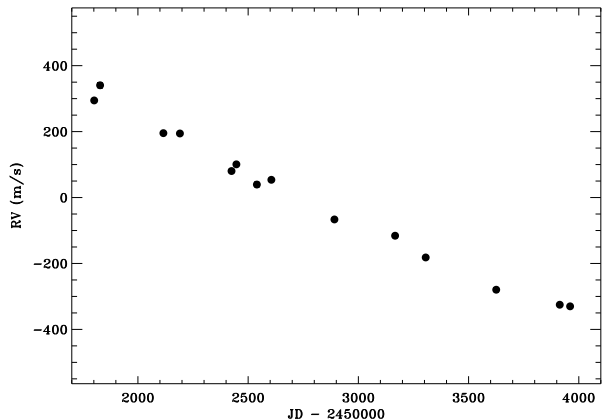}
\caption{Radial velocities curve of one of the stars showing a clear linear trend, with a marginal
indication for curvature in the last season.}
\label{f:trend}  
\end{figure}

In most cases the trends are due to new low mass, possibly substellar
companions orbiting one of the components. 
One example is shown in Fig. \ref{f:trend}. In two cases, the trends 
show highly significant curvature
and the radial velocity curves are compatible with massive planets with
period longer than 7-10 yr.
The continuation of the radial velocity monitoring will reveal the period and 
nature of these objects.

\begin{figure}
\centering
\includegraphics[height=3.85cm,angle=90]{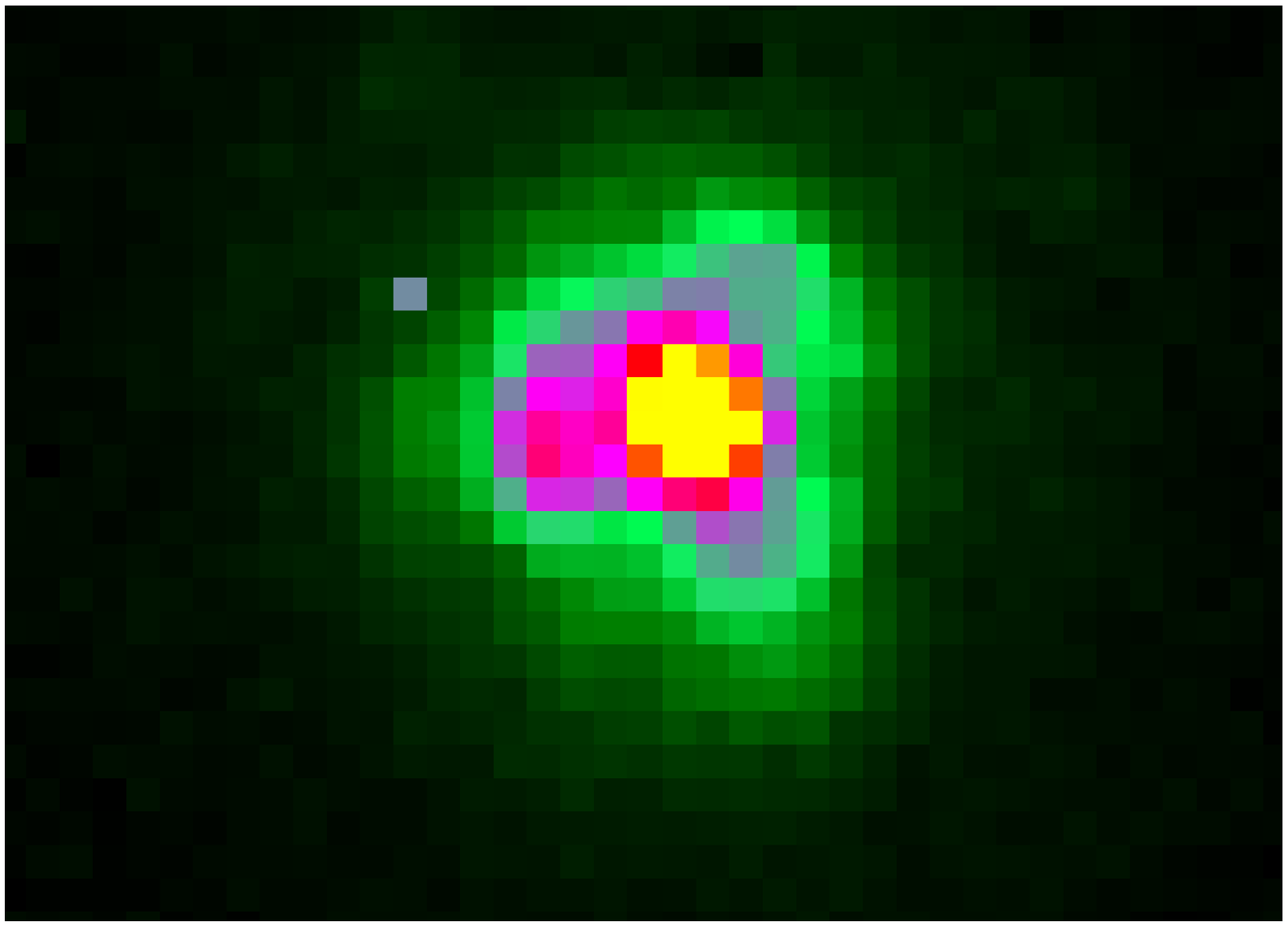}
\includegraphics[height=3.85cm,angle=90]{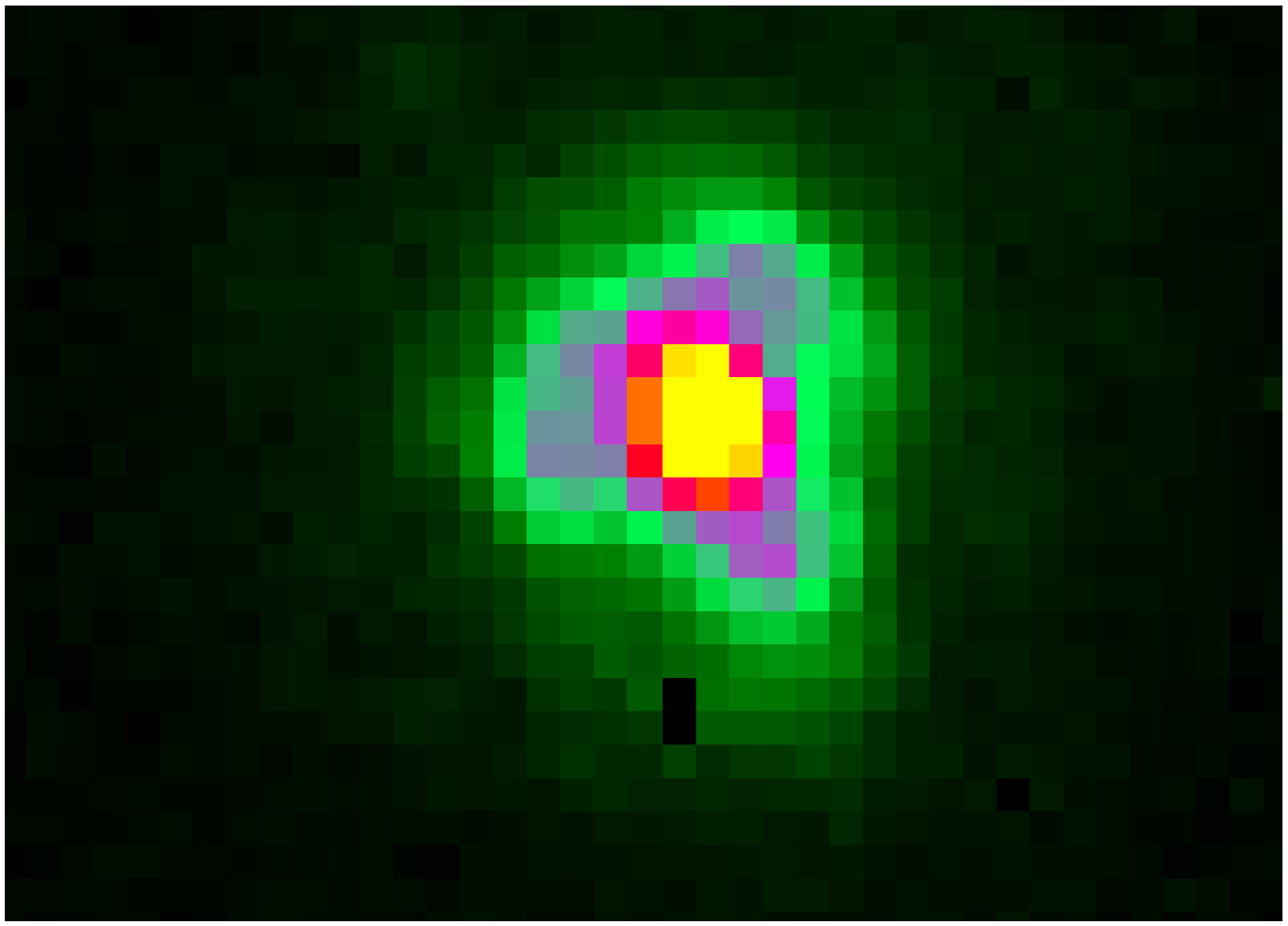}
\includegraphics[height=3.85cm,angle=90]{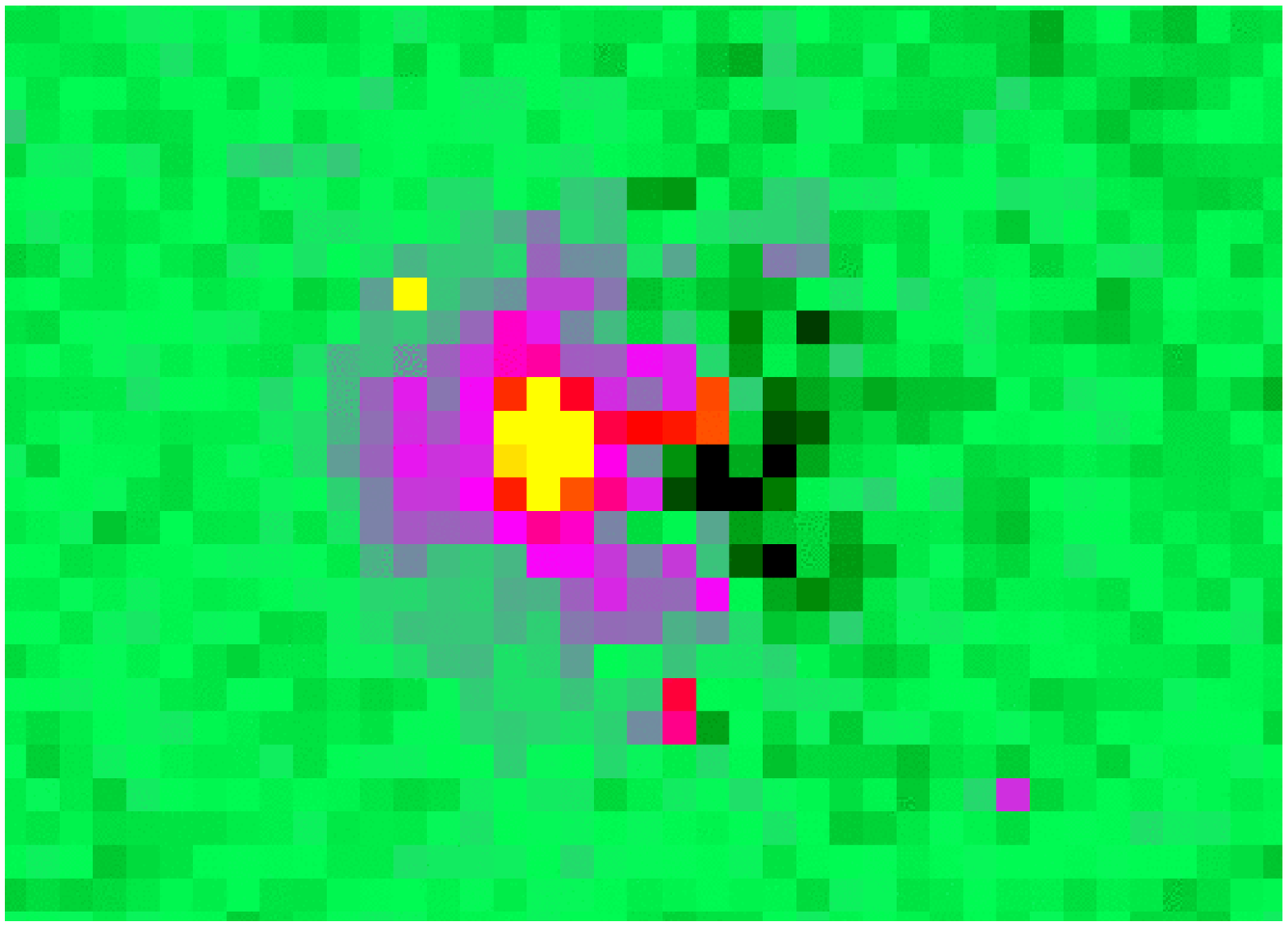}
\caption{Adaptive optics identification of a close companion around a star with
RV linear trend (Fig.~\ref{f:trend}): left panel: image the star with RV trend, central panel: image of the
wide companion; right panel: difference between the two images. PSF artefacts were removed
fairly well, allowing the identification of a close companion at 0.2 arcsec from the star. 
This is probably responsible for the observed RV trend. }
\label{f:adopt}       
\end{figure}

We recently started an adaptive optics program to identify the companions 
of stars with long term trends using AdOpt@TNG (\cite{adopt}). 
Preliminary results for one object are shown in Fig.~\ref{f:adopt}. 
The direct identification of substellar objects as companions of stars for
which age and chemical composition can be derived would play a relevant role
in the calibration of models of substellar objects.
It  also allows us a better characterization of the orbits and mass
ratios of the systems we are monitoring. This point is relevant for the
studies of the frequency and properties of planets in binaries as a function
of the binary separation or effective gravitational influence.

Finally, we also detected a few new spectroscopic binaries
among the components of the wide binaries. These systems are then composed
by at least three components. Some of these systems are presented in
\cite{garching}.

\section{Line bisectors: a tool to study stellar activity and contamination}
\label{s:bis}

The relevance of activity jitter for the interpretation of the RV
data prompted us to develop a tool to measure and possibly to correct for 
its effect.
The differential RV variations induced by stellar activity are due to changes 
in the profile of spectral lines caused by the presence of spots and/or the 
alteration of the granulation pattern in active regions.
The activity jitter of a star may be predicted by means of statistical 
relations from its chromospheric emission, rotational velocity or amplitude 
of photometric variations (\cite{saar98}; \cite{paulson04}; \cite{wright05}). 
Simultaneous determination of RV, chromospheric emission and/or photometry 
is even more powerful in disentangling the origin of the observed RV 
variations Keplerian vs. stellar activity.
The measurement of the line profile alterations on the same spectra 
(ideally on the same spectral lines) represents a direct measurement
of the activity jitter.
The existence of a correlation between the variations
of the RV and those of the line profile is a strong indication for 
non-Keplerian origin for the observed RV variations.

The study of line profile as a tool to disentangle Keplerian
motion and activity jitter is usually performed 
using a few well isolated lines on high S/N spectra (see e.g. \cite{hatzes})
or by combining the cross-correlation profiles of many spectral lines
at moderate S/N ratios with a suitable template mask 
(see e.g., \cite{queloz01}).

In our case, we followed the latter approach, but we had to handle
the complication of having the iodine lines superimposed to the stellar
spectra. On the other hand, these lines offer the opportunity to improve the
wavelength calibration of the spectra, required for accurate estimates 
of the line bisectors.
The iodine lines were removed by means of a suitable spectrum 
of a fast rotating early type star with the iodine cell in the optical path.
The procedure is described in detail in \cite{aldo}.

\begin{figure}
\centering
\includegraphics[height=11cm]{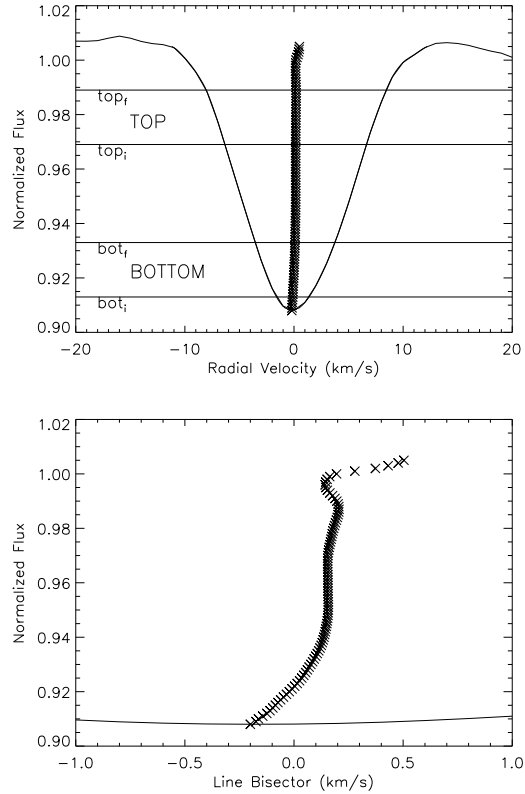}
\caption{Spectrum of HD 166435. In the top panel we show the normalized 
cross correlation profile, the line bisector, the top and bottom zones 
(both with 
$\Delta F = 0.02$; $\Delta F =\rm top_{f}~ -~ top_{i} = bot_{f} ~-~ bot_{i})$. 
In the bottom panel we show a zoom of the profile with the  RV scale 
increased to better display the asymmetries of the line bisector. 
From \cite{aldo}.}
\label{f:bvs}       
\end{figure}

The bisector of an absorption line is the middle point of the horizontal 
segment connecting points on the left and right sides of the profile with 
the same flux level. The line bisector is obtained by combining bisector 
points ranging from the core toward the wings of the line.
To quantify the asymmetry of the spectral lines and look for correlation 
with RV it is useful to introduce the bisector velocity span 
(hereafter BVS, \cite{toner}). This is determined by 
considering a top zone near 
the wings and a bottom zone close to the core of the lines, which represent 
interesting regions to study the velocity given by the bisector 
(see Fig.~\ref{f:bvs}). 
The difference of the average values of velocities in the top and bottom 
zones, $V_{\rm T}$ and $V_{\rm B}$ respectively, determine the bisector 
velocity span.

\begin{figure}
\centering
\includegraphics[height=8.0cm,angle=90]{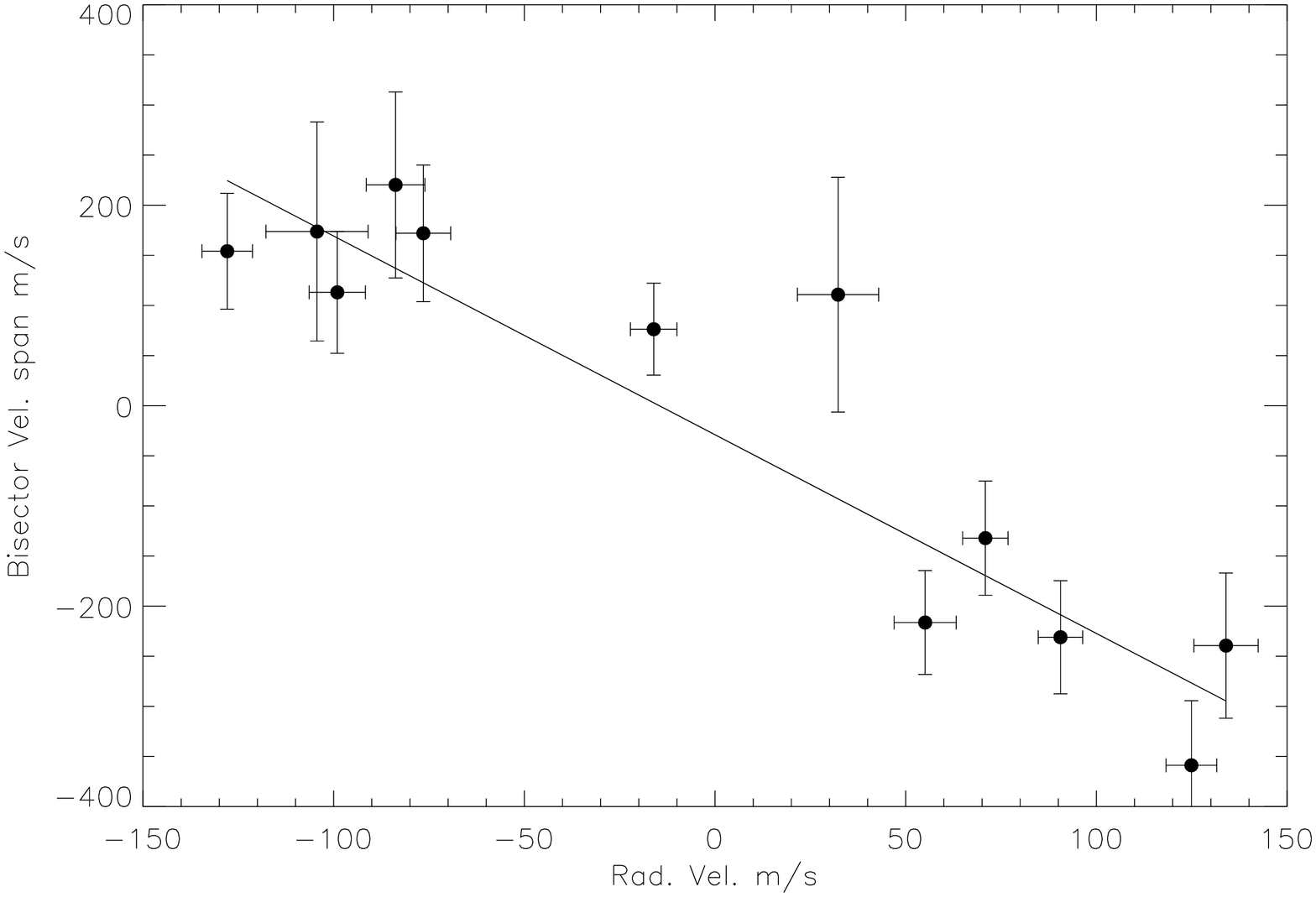}
\caption{Radial velocity - line bisector correlation for the active star 
HD~166435}
\label{f:hd166435}       
\end{figure}

The  star HD 166435 shows evidence of RV variations, 
photometric variability and magnetic activity. Furthermore, previous analysis 
of the variation of the line bisectors revealed a correlation between 
RV and line bisector orientation (\cite{queloz01}).
It was used to test our procedure. As shown in Fig.~\ref{f:hd166435}, 
there is a clear anti-correlation between radial velocity and BVS variations.

The study of line shape is relevant for our program also as a diagnostic for
the contamination of the spectra by the wide companion.
Contaminated spectra are not easy to handle when analyzing the radial
velocity curve. In fact, the internal radial velocity errors are estimates
from the scatter of individual chunks on which the spectrum is modeled
separately. In case of contamination, all the chunks deviate systematically
by a similar amount (our pairs are always formed by similar stars) and 
then the radial velocity shift might largely exceed the internal errors, 
causing a spurious  but formally  highly significant variability.

\begin{figure}
\centering
\includegraphics[height=8.0cm,angle=90]{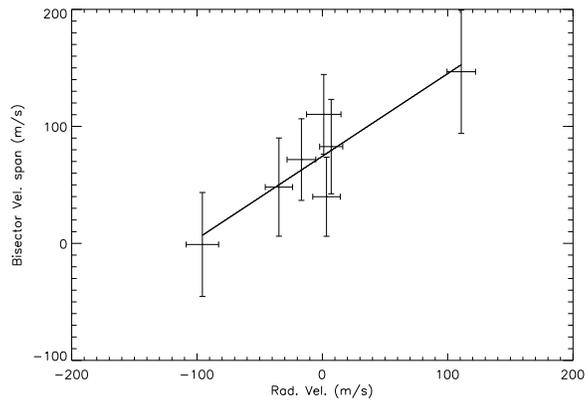}
\caption{Radial velocity - line bisector correlation for HD~8071B.
This is likely due to the comtamination by the companion HD~8071A}
\label{f:hd8071}       
\end{figure}

In the case of contamination, we observe a positive correlation between
the bisector velocity span and the radial velocity.
The worst case of contamination in our sample occurs for HD~8071B
(see Fig.~\ref{f:hd8071}).
This pair is one of the closest  (separation 
2.1 arcsec). Furthermore, HD 8071~A is itself a single-lined
spectroscopic binary with a RV semi-amplitude of about 7 km/s. This causes
significant spectra-to-spectra variations of the 
contamination both in amplitude
(because of the variable observing conditions) and wavelength
(because of the orbital motion of HD~8071A).

\section{Upper limits on planetary companions}
\label{s:limits}

While no confirmed planet detection emerged up to now from our survey,
a detailed analysis of the negative results would allow to constrain
the frequency of planets in binary systems. Since we are focusing on a
specific type of binaries, wide binaries with similar components
at intermdiate separations (a few hundreds AU), 
such a study is complementary to other studies of planets in binaries.

To this aim, we derived upper limits on the planetary companions 
still compatible with the observations. 
Our method, a Montecarlo simulation based on the evaluation of the 
excess of radial velocity variability caused
by the presence of hypothetical planets, allows us a complete exploration
of the possible orbital parameters for eccentric orbits (the
real case, since most of the known planets are in eccentric orbits).
Our approach is described in detail in \cite{hd219542}.

\begin{figure}
\centering
\includegraphics[height=5.5cm,angle=90]{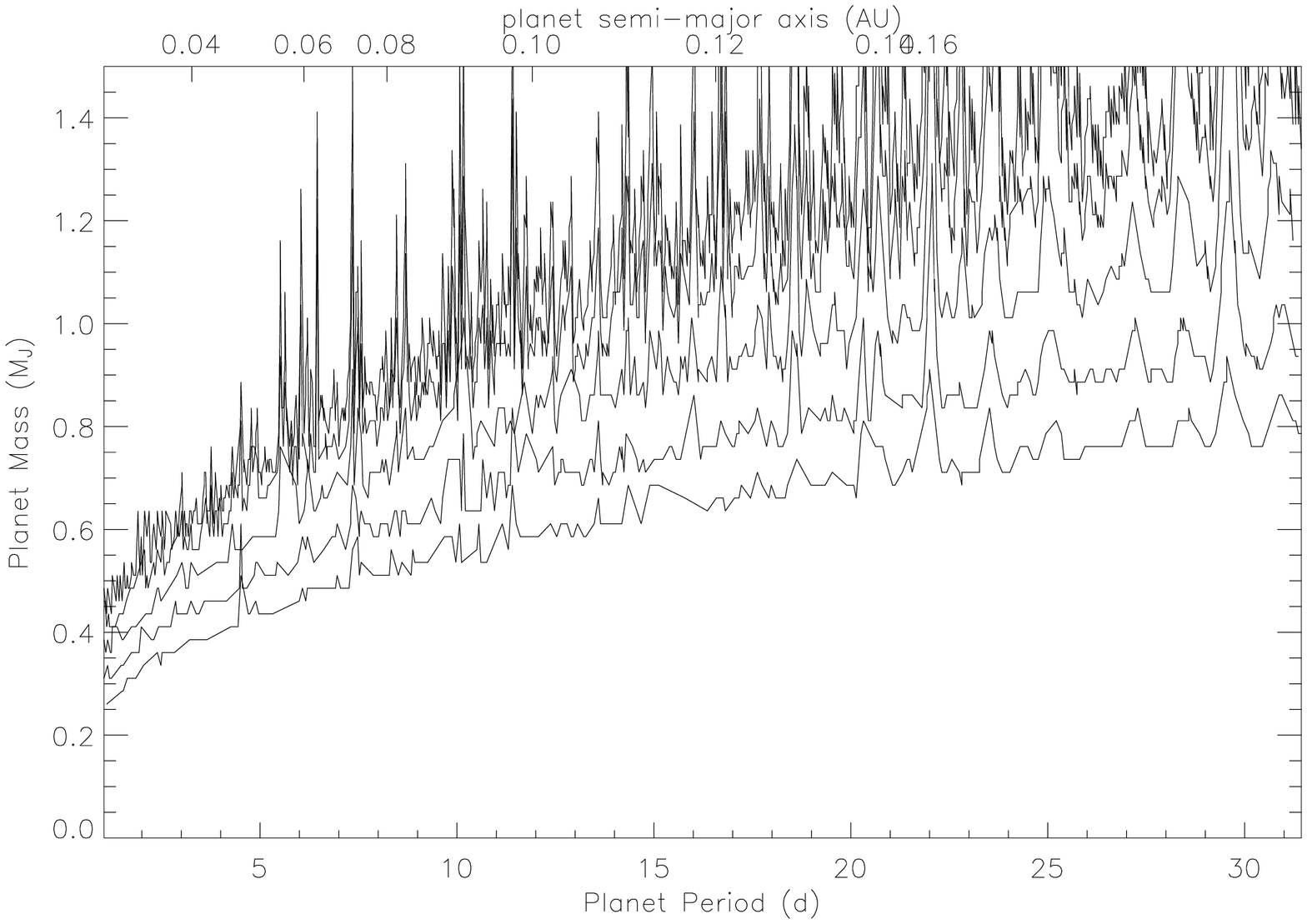}
\includegraphics[height=5.5cm,angle=90]{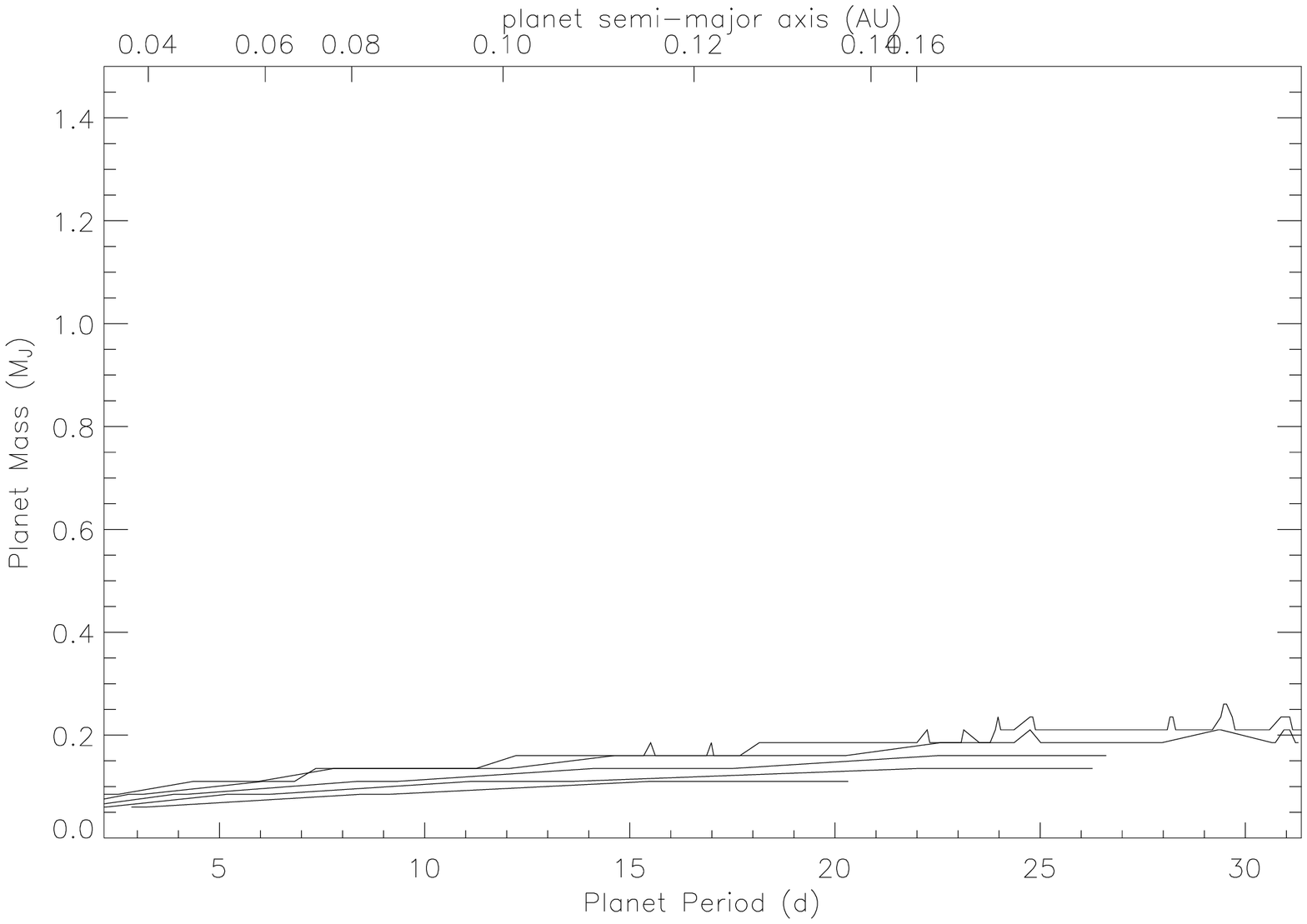}
\includegraphics[height=5.5cm,angle=90]{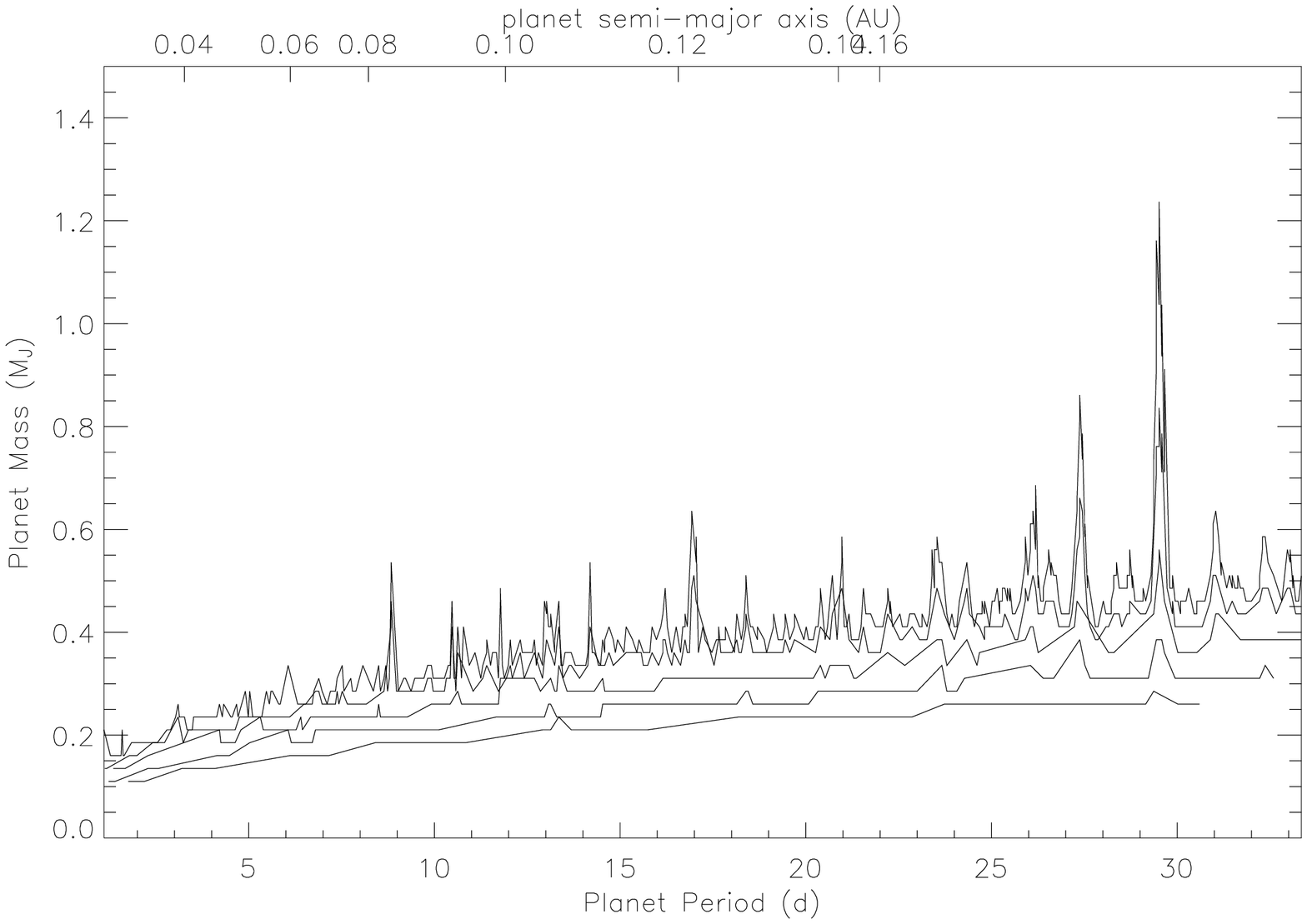}
\includegraphics[height=5.5cm,angle=90]{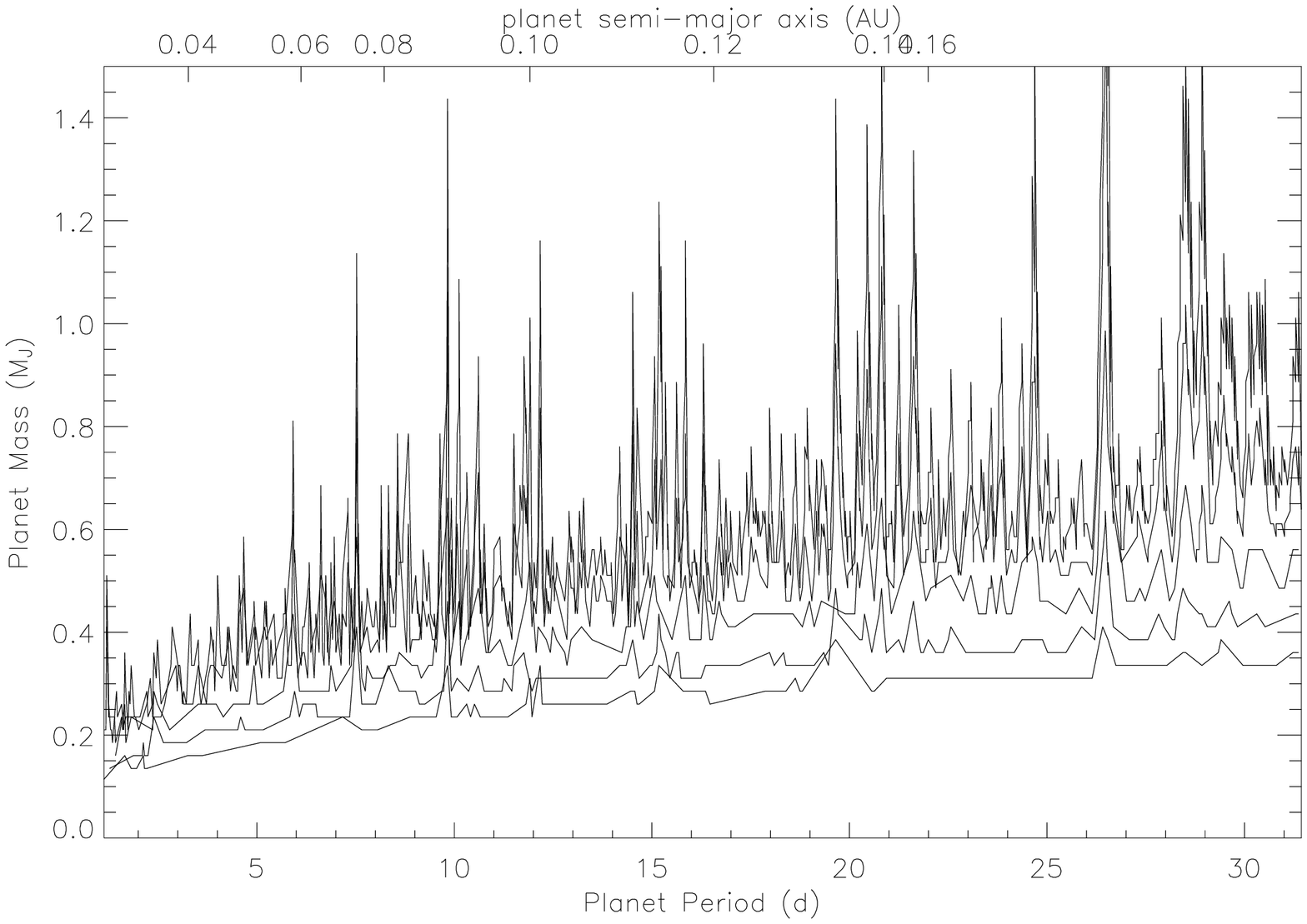}

\caption{Upper limits on planetary companion on short-period circular orbit 
for four stars representative of our sample. 
The different lines refer to the exclusion limits for  (from top to botton) 
95\%, 90\%, 75\%,  50\%, and 25\% of the planets. 
For the star on the upper-left 
corner planet detectability is strongly limited by stellar activity. 
The star in
the upper-right corner is the one with the best limits, thanks to the 
low dispersion
of RVs and the large number of measurements. The behaviour of the 
other two stars is more typical for our survey. The 'noisy' run of 
exclusion limits with period for the star in the lower-right corner
is due to the small number of measurements. }
\label{f:limits_close}       % Give a unique label
\end{figure}

\begin{figure}
\centering
\includegraphics[height=5.5cm,angle=90]{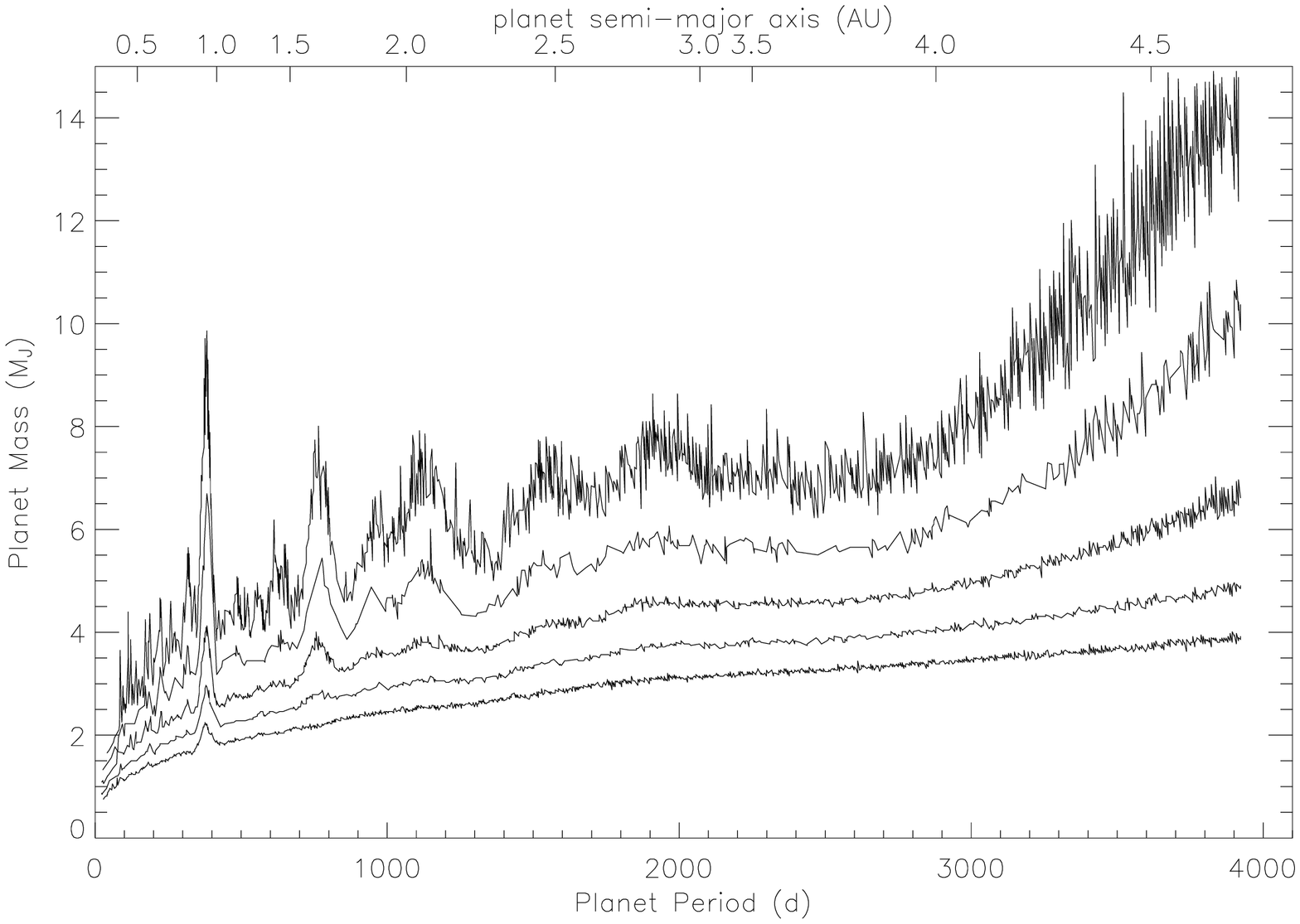}
\includegraphics[height=5.5cm,angle=90]{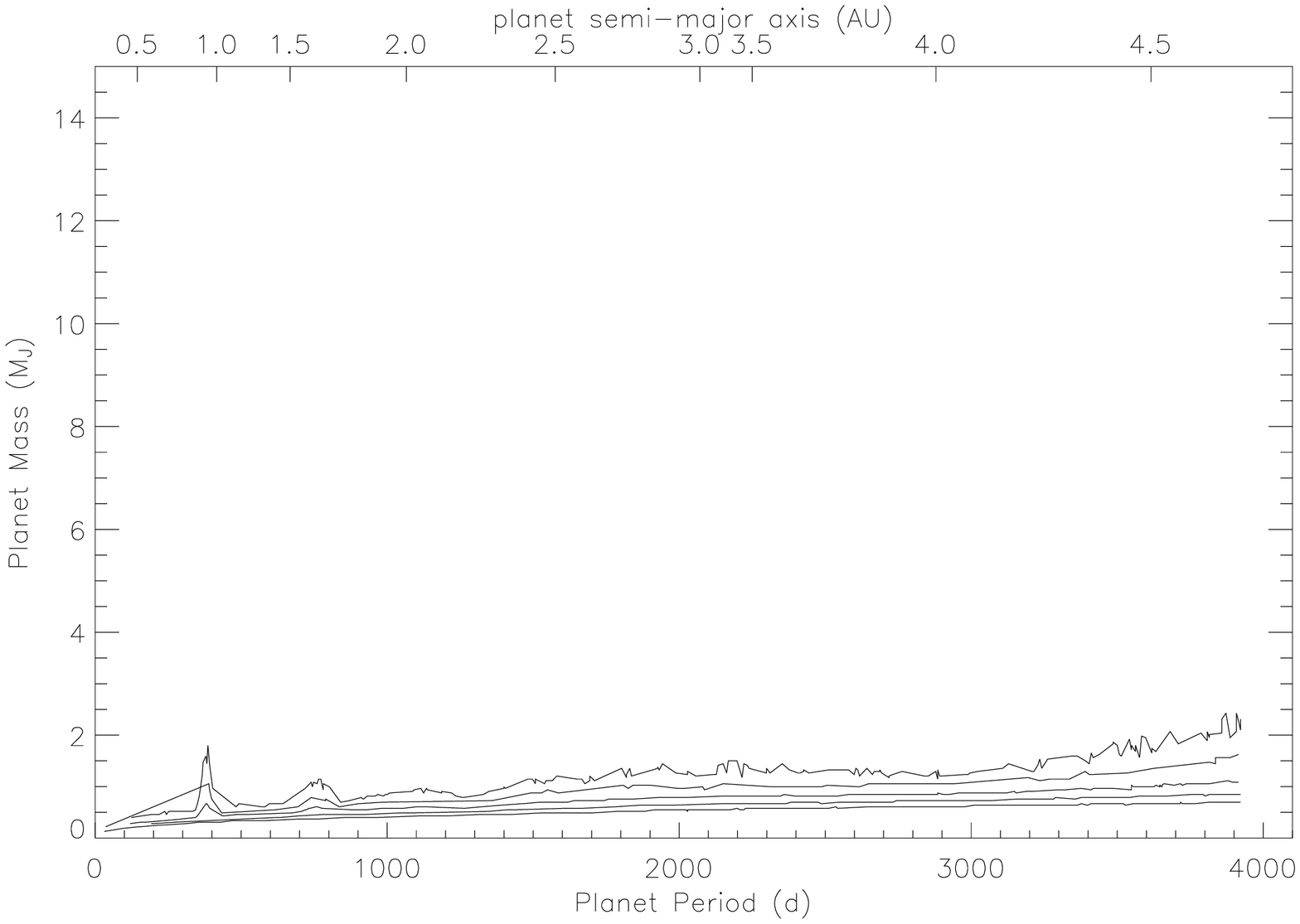}
\includegraphics[height=5.5cm,angle=90]{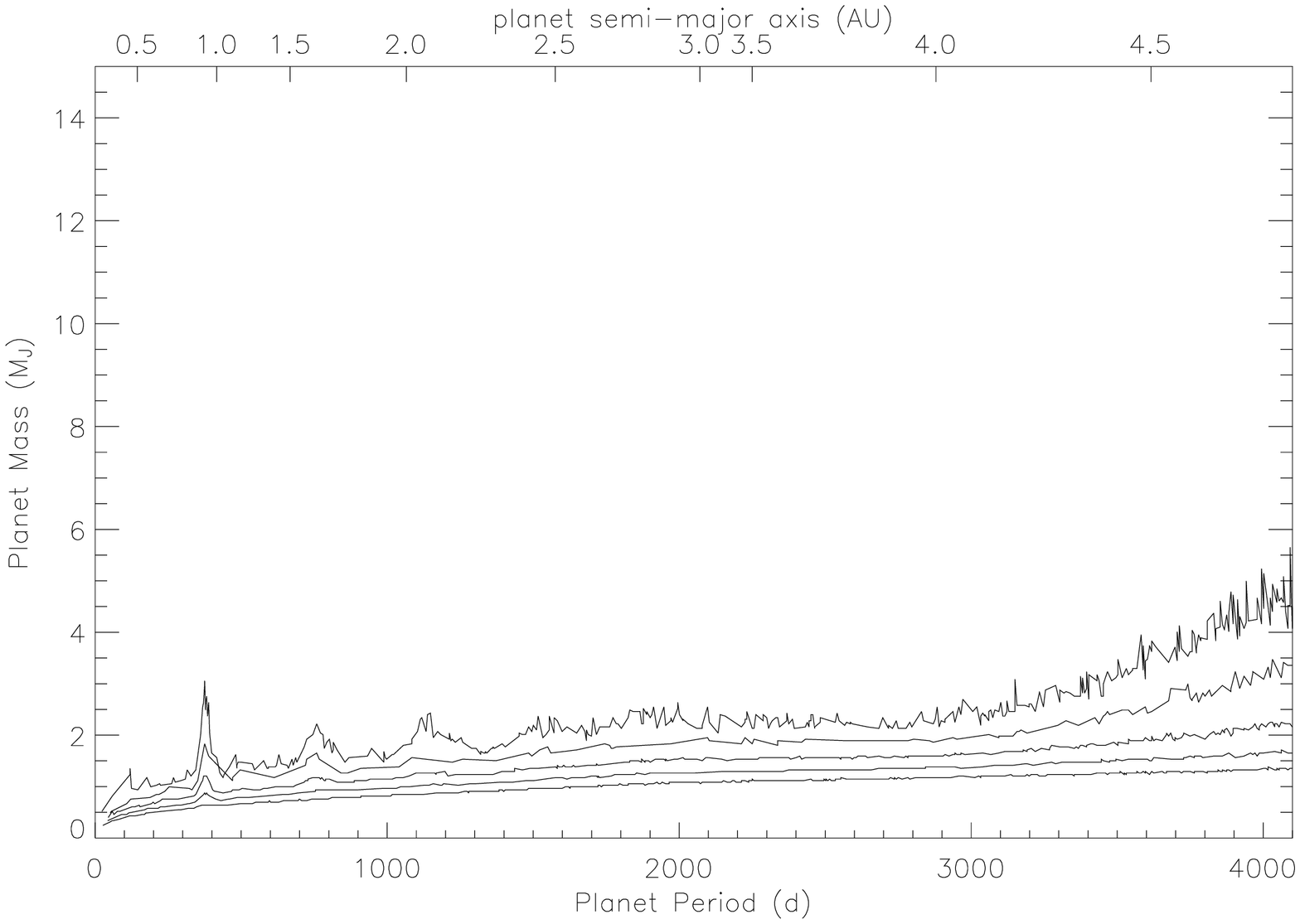}
\includegraphics[height=5.5cm,angle=90]{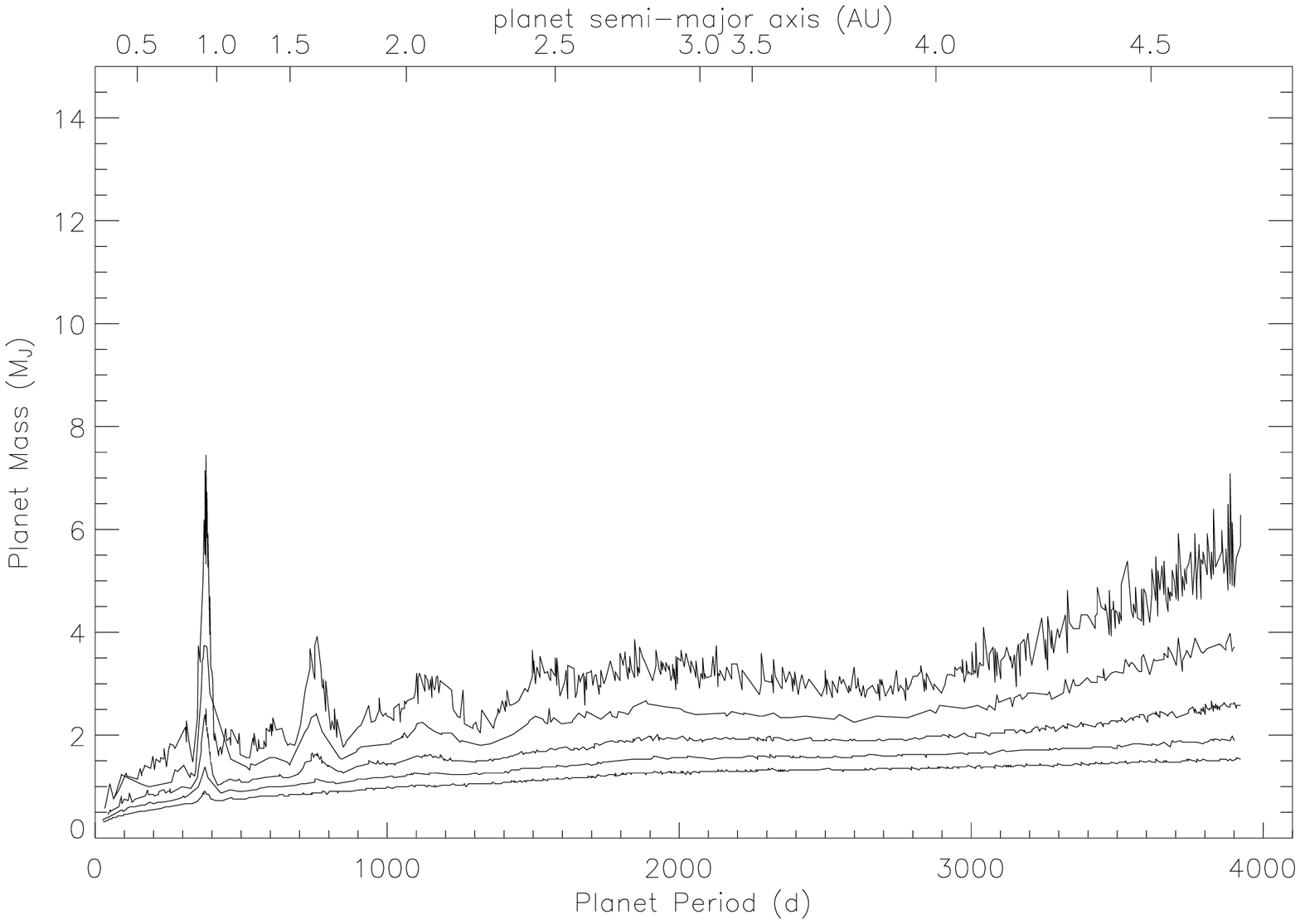}

\caption{Upper limits on planetary companion on long-period eccentric orbit 
for the same four stars shown in Fig.~\ref{f:limits_close}.}
\label{f:limits_ecc}       
\end{figure}

\begin{figure}
\centering
\includegraphics[height=9.0cm,angle=90]{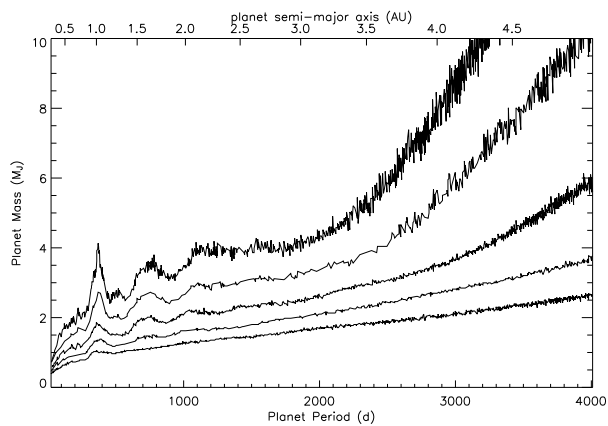}
\caption{Summary of estimates of exclusion/compatibility of planets 
in the SARG sample with
current data for the stars with
at least 10 observations. For each period,
the mass corresponding to the exclusion of (from top to botton)  95\%, 90\%,
75\%,  50\%, and 25\% of the planets (taking into account planet eccentricity) 
is shown. 
The results of individual stars were averaged to produce the plot.} 
\label{f:summary}       
\end{figure}

Fig.~\ref{f:limits_close} shows the upper limits on planetary companion on 
short-period circular orbit for four stars representative of our sample.
Fig.~\ref{f:limits_ecc} shows the limits for long period planets 
with eccentricities as large as 0.95.
The average limits for the whole sample are shown in Fig.~\ref{f:summary}.

\section{On the frequency of planets in binary systems}
\label{s:freq}

The lack of planets up to now in the SARG sample appears as an indication
for a lower frequency of planets in the kind of binary systems we are
surveying.
Since our sample includes only binaries, a reference sample is needed
for a full statistical evaluation.
A useful comparison sample is represented by the 'Uniform Detectability'
sample identified by \cite{fv05}.

The Uniform Detectability (UD) sample has been built from the full 
target lists of Lick, Keck and Anglo Australian Surveys (1330 stars),
satisfying the requirement of completeness for detections of planets
with velocity amplitudes K$>$30 m/s and orbital periods shorter than 4 years.
Stars that were added after a planet was discovered by other groups were not 
included in the sample.
However, stars independently present in one of these surveys were considered 
even if a planet was detected first by another group.
Only planets with K$>$ 30 m/s and orbital periods shorter than 4 years 
were considered for the study of planet frequency. 
This corresponds to Saturn-mass planets for the shortest periods and 
Jupiter-mass planets for 4 year orbits.

The UD sample is biased against binaries, as the stars with companions closer
than 2 arcsec known at the time of the target selection were excluded.
Bonavita \& Desidera (\cite{bd06}) performed a detailed literature 
search for binarity of the 850
stars in the UD sample, resulting in  202 binary stars in the sample.
For some of them, only long term radial velocity and astrometric trends
are available.

15 of the binaries in the UD sample have planets, 
so the global frequency of planets in the UD 
binary sample is 7.4\%.
If we consider the single-stars sub-sample, we found that 5.3\% of UD 
single stars  have planets (see Table 2).
The two frequencies are compatible within their errors. The slightly 
higher value of the 
global frequency in the binary sub-sample is probably due to higher 
completeness 
level of binary census in stars with planet.

Incompleteness effects are unlikely to deeply modify this picture. 
Even  assuming that the frequency 
of binaries in the sample is that found by \cite{duq91} (an upper 
limit because of the exclusion
of binaries with separation less than 2 arcsec) and that all the 
companions of planet hosts
have been already identified, it can be seen that the global frequency 
of planets in binaries 
can not be lower by more than a factor of three  compared to that of 
single stars.

The rather large sample size allows us we make some sub-samples with 
different values of 
critical semiaxis for dynamical stability of planets ($a_{crit}$, see 
\cite{holman} and Sect.~\ref{s:bin}).
All the stars with RV and/or astrometric trend are included in the 
closest bin, as it is likely that 
the companion responsible of the trend is at small separation.

\begin{table}
\begin{center}
\begin{tabular}{|c|c|c|c|}
\hline
  $a_{crit}$  & $N_{stars}$ & $N_{planets}$ & $\frac{N_{planets}}{N_{stars}}$\\
\hline
     20 AU   & 89&  2&0.022$\pm$0.018\\
20 - 50 AU   & 18&  2&0.111$\pm$0.105\\
50 - 100 AU  & 24&  2&0.083$\pm$0.076\\
100 - 250 AU & 26&  4&0.154$\pm$0.107\\
$>$ 250 AU   & 45&  5&0.111$\pm$0.066\\
\hline
\hline
UD Singles sub-sample&647& 34&0.053$\pm$0.011\\
Entire UD binary sub-sample&202& 15&0.074$\pm$0.024\\
\hline
\end{tabular}
\caption{\footnotesize Frequency of planets in binaries with different 
values of  $a_{crit}$. From \cite{bd06}}
\end{center}
\label{t:a_bin}
\end{table}

We found that there is no significant dependence of the frequency on 
$a_{crit}$  except for companion with 
$a_{crit}$ less than 20 AU (that corresponds to a separation
 $<$ 50-100 AU, depending on 
the mass-ratio of the components).
Considering also the similitude of the mass and period distribution of planets 
orbiting single stars and components of wide binaries 
(see \cite{db06} and Sect.~\ref{s:bin}),
we then conclude that a wide companion plays a marginal role on the 
formation and evolution of giant planets. 

For the planets in tight binaries, the results are more intriguing.
On one hand, there are indications that the properties of planets in 
tight binaries are 
significantly different from those of exoplanets orbiting wide binaries 
or single stars (see \cite{db06} and Sect.~\ref{s:bin}).
On the other hand, the frequency of planets in close binaries appears 
to be lower 
than that of planets orbiting single stars and components of 
wide binaries.

The frequency of planets in close binaries can be used to further 
investigate how these
planets formed and the origin of their anomalous properties.
Indeed, \cite{pfhal06} showed that the knowledge of the value of the 
frequency of 
planets in close binaries\footnote{Defined as those binaries with 
semi-major axis less 
than 50 AU} should allow to disentangle between two alternative 
formation scenarii. 
A low frequency (less than 0.1\% but with an uncertainty of about 
one order of magnitude, so 
they consider 1\% as a limit-value) 
would be compatible with dynamical interactions that cause the 
formation of the tight binary after planet formation.
While not fully conclusive because of the poor statistics, our results 
suggests that frequency of planets in close binaries probably is not as 
low as required to explain their presence only as the results of 
modifications of the binary orbit after the planet formation.
Therefore, it appears that  planets do form in tight binaries 
(separations of the order of 20 AU or even less) in spite of the 
strong gravitational interactions that might work against.

\begin{figure}
\centering
\includegraphics[height=6.5cm]{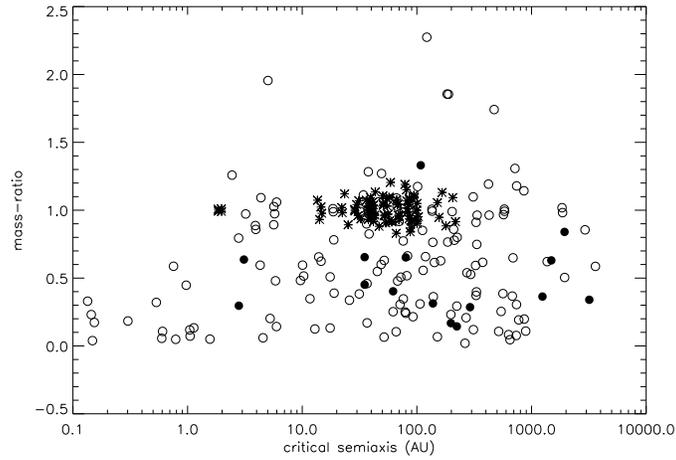}
\caption{Critical semiaxis for dynamical stability for the binaries in the UD sample 
(filled circles: stars with planets; empty circles: stars without planets) and in the sample
of the SARG planet search (asterisks).}
\label{f:ud_bin}       
\end{figure}

However, crucial issues still need clarification.
There are some hints that the run frequency of planets is not 
characterized by a continuous decrease when moving to smaller separation:
in the full list of planets in binaries by  \cite{db06} there is
only one planet with  critical semimajor axis 
for dynamical stability in the range 10 - 30 AU, while there are 
5 planets with $a_{crit}$ less than 
10 AU and 4 planets with 30 $<$ $a_{crit}$ $<$ 50 AU.
This suggests a bimodal distribution of planet frequency, with 
a secondary maximum at $a_{crit} \sim 3-5$  AU, but the analysis of the 
UD sample does not allow us to confirm it because of the small number
of binaries with 10 $<$ $a_{crit}$ $<$ 30 AU and the lack  
of binary characterization (orbital parameters, 
mass ratio) for the stars with only RV and/or astrometric trends.

The targets of the SARG planet search are crossing this range of 
separation (see Fig.~\ref{f:ud_bin}), and therefore
the completion of the survey, coupled with an estimate of planet 
detectability homogeneous
with that of comparison samples will allow us to better address this issue.
The current lack of planets in the SARG survey might suggest a relevant 
role of the binary
mass ratio in the occurrence of planets.
A complementary very important approach is represented by  a detailed 
characterization of the binaries in current samples of RV surveys 
(complete detection of binaries and, when possible, full determination of the
orbital elements).
The availability of a larger and more complete sample will allow us 
to better understand the behaviour 
of the planet frequency in binaries and, at the same time, to disentangle 
the questions about the formation 
of planets in these peculiar environments and especially  about the 
formation mechanisms and the different 
characteristics of the planets in tight binaries.

% BibTeX users please use
% \bibliographystyle{}
% \bibliography{}
%
% Non-BibTeX users please follow the syntax
% the syntax of "referenc.tex" for your own citations

%%%%%%%%%%%%%%%%%%%%%%%% referenc.tex %%%%%%%%%%%%%%%%%%%%%%%%%%%%%%
% sample references
% "physics"
%
% Use this file as a template for your own input.
%
%%%%%%%%%%%%%%%%%%%%%%%% Springer-Verlag %%%%%%%%%%%%%%%%%%%%%%%%%%

%
% BibTeX users please use
% \bibliographystyle{}
% \bibliography{}

\begin{thebibliography}{99.}
% Use the following syntax and markup for your references
%
% Monographs

\bibitem{bakos} G.A. Bakos, R.W. Noyes, G. Kovacs et al: ApJ, in press (astro-ph 0609369) 
\bibitem{bd06} M. Bonavita, S. Desidera: A\&A submitted (2007)
\bibitem{butler06} R.P. Butler, J.T. Wright, G.W. Marcy, et al. ApJ  \textbf{4646}, 505 (2006)
\bibitem{adopt} M. Cecconi, A. Ghedina, P. Bagnara et al.: Proceedings of the SPIE, Volume 6272, p. 77  (2006)
\bibitem{db06} S. Desidera, M. Barbieri: A\&A in press (2007)
\bibitem{hd219542} S. Desidera, R. Gratton, M. Endl  et al.: A\&A \textbf{405}, 207 (2003)
\bibitem{chem2} S. Desidera, R. Gratton, S. Scuderi  et al.: A\&A \textbf{420}, 683 (2004)
\bibitem{letter} S. Desidera, R. Gratton, M. Endl  et al.: A\&A \textbf{420}, L27 (2004)
\bibitem{garching} S. Desidera, R. Gratton, R.U. Claudi: in
                \textit{Proc. of ESO Workshop on Multiple Stars Across HR Diagram}, in press (2006)
\bibitem{chem3} S. Desidera, R. Gratton, S. Lucatello, R.U. Claudi.: A\&A \textbf{454}, 581 (2006)
\bibitem{bs} S. Desidera, R. Gratton, S. Lucatello, M. Endl, S. Udry: A\&A, in press (2006)
\bibitem{duq91} A. Duquennoy \& M. Mayor: A\&A  \textbf{248}, 485 (1991)
\bibitem{egg04} A. Eggenberger, M. Mayor M. S. Udry: A\&A \textbf{417}, 353 (2004)
\bibitem{egg06} A. Eggenberger, S. Udry, M. Mayor M. et al.: in
                \textit{Proc. of ESO Workshop on Multiple Stars Across HR Diagram}, in press (2006) 
\bibitem{austral} M. Endl, M. K\"urster, S. Els: A\&A, \textbf{362}, 585 (2000)
\bibitem{fv05} D. Fischer \& J. Valenti: ApJ  \textbf{622}, 1102 (2005)
\bibitem{gonzalez} G. Gonzalez: MNRAS \textbf{285}, 403 (1997)
\bibitem{sarg} R.G. Gratton, G. Bonanno, P. Bruno, et al.:  Exp. Astron. \textbf{12}, 107 (2001)
\bibitem{hatzes} A.P. Hatzes, W.D. Cochran, E.J. Bakker: ApJ \textbf{508}, 380 (1998) 
\bibitem{hatzes05} A.P. Hatzes  \& G. Wuchterl:  Nat. \textbf{436}, 182 (2005) 
\bibitem{heiter} U. Heiter, R.E. Luck: AJ, \textbf{126}, 201 (2003)
\bibitem{holman} M.J. Holman, P.A.. Wiegert: AJ, \textbf{117}, 621 (2001)
\bibitem{laws} C. Laws, G. Gonzalez: ApJ, \textbf{553}, 405 (2001)
\bibitem{aldo} A.F. Martinez Fiorenzano, R. Gratton, S. Desidera, R. Cosentino, M. Endl: A\&A, \textbf{442}, 775 (2005)
\bibitem{marzari02} F. Marzari  \& S.J. Weidenschilling:  Icarus \textbf{156}, 570 (2002)
\bibitem{murray} N. Murray, B. Chaboyer, P. Arras, B. Hansen, \& R.W. Noyes: ApJ \textbf{555}, 801 (2001)
\bibitem{nordstrom} B. Nordstrom, M. Mayor, J. Andersen et al.: A\&A \textbf{418}, 989 (2004)
\bibitem{paulson04} D.B. Paulson, S.H. Saar, W.D. Cochran, G.W. Henry: AJ  \textbf{127}, 1644 (2004)
\bibitem{pfhal06} E. Pfhal \& M. Mutherspaugh: ApJ \textbf{652}, 1694 (2006)
\bibitem{preston} G.W. Preston \& C. Sneden:  AJ \textbf{120}, 1014 (2000)
\bibitem{queloz01} D. Queloz, G.W. Henry, J.P. Sivan,  et al.:  A\&A \textbf{379}, 279 (2001)
\bibitem{saar98} S.H. Saar, R.P. Butler \& G.W. Marcy:  ApJ, \textbf{498}, L153 (1998)
\bibitem{santos04} N.C. Santos, G. Israelian , M. Mayor: A\&A \textbf{415}, 1153 (2004)
\bibitem{soder} S. Soderhjelm: A\&A \textbf{341}, 121 (1999)
\bibitem{takeda06} G. Takeda, E.B. Ford, A. Sills, A. et al.: 
      ApJS, in press (2007) (astro-ph 0607235)
\bibitem{takeda} Y. Takeda: PASJ \textbf{57}, 83 (2005)
\bibitem{taylor05} B.J. Taylor: ApJS \textbf{161}, 444 (2005)
\bibitem{tok06} A.A. Tokovinin, S. Thomas, M. Sterzik, S. Udry, S. 
        A\&A   \textbf{450}, 681 (2006)
\bibitem{toner} C.G. Toner  \& D.F. Gray: ApJ, \textbf{334}, 1008 (1988)
\bibitem{vf05} J. Valenti \& D. Fischer: ApJS  \textbf{159}, 141 (2005)
\bibitem{vanwinckel} H. Van Winckel: ARA\&A, \textbf{41}, 391 (2003)
\bibitem{winnick02} R.A. Winnick, P. Demarque, S. Basu, D.B. Guenther:  ApJ \textbf{576}, 1075 (2002)
\bibitem{wright05} J.T. Wright:  PASP \textbf{117}, 657 (2005)
\bibitem{zucker02} S. Zucker, T. Mazeh:  ApJ \textbf{568}, L113 (2002)



%\bibitem{monograph} H. Ibach, H. L\"uth: \textit{Solid-State
%Physics}, 2nd edn (Springer, Berlin Heidelberg New York 1996) pp 45--56

% Contributed Works

%\bibitem{contribution} D.M. MacKay: Visual stability and voluntary eye
%movements. In: \textit{Handbook of Sensory Physiology}, vol 3, ed by R.
%Jung, D.M. MacKay (Springer, Berlin Heidelberg New York 1973) pp
%307--331

% Journal
%\bibitem{journal} S. Preuss, A. Demchuk Jr, M. Stuke et al: Appl. Phys.
%A \textbf{61}, 33 (1995)

% Theses
%\bibitem{thesis} D.W.  Ross: Lysosomes and storage diseases. MA
%Thesis, Columbia University, New York (1977)

\end{thebibliography}

%%%%%%%%%%%%%%%%%%%%%%%%%%%%%%%%%%%%%%%%%%%%%%%%%%%%%%%%%%%%%%%%%%%%%%  }

%%%%%%%%%%%%%%%%%%%%%%%%%%%%%%%%%%%%%%%%%%%%%%%%%%%%%%%%%%%%%%%%%%%%%%

\printindex
\end{document}